\documentclass[12pt]{article}
\pdfoutput=1
\usepackage{pstricks}
\usepackage{color}
\usepackage{amssymb,amsmath,bm,bbm}
\usepackage{epsf}
\usepackage{epsfig}
\usepackage{afterpage}
\usepackage{longtable}
\usepackage{cite}
\usepackage{latexsym,mathrsfs,dsfont}
\usepackage{graphics}
\usepackage{url}
\usepackage{paralist}
\usepackage{bbold}

\newcommand{\tev}{\, {\rm TeV}}
\newcommand{\gev}{\, {\rm GeV}}
\newcommand{\mev}{\, {\rm MeV}}

\newcommand{\vcb}{|V_{cb}|}

\newcommand{\vub}{|V_{ub}|}

\newcommand{\beq}{\begin{equation}}
\newcommand{\eeq}{\end{equation}}
\newcommand{\be}{\begin{equation}}
\newcommand{\ee}{\end{equation}}
\newcommand{\bi}{\begin{itemize}}
\newcommand{\ei}{\end{itemize}}
\newcommand{\ba}{\begin{array}}
\newcommand{\ea}{\end{array}}
\newcommand{\beqa}{\begin{eqnarray}}
\newcommand{\eeqa}{\end{eqnarray}}
\newcommand{\bea}{\begin{eqnarray}}
\newcommand{\eea}{\end{eqnarray}}
\newcommand{\beqn}{\begin{eqnarray}}
\newcommand{\eeqn}{\end{eqnarray}}

\newcommand{\D}{\Delta}

\newcommand{\eps}{\epsilon}

\definecolor{red}{cmyk}{0,1,1,0.4}

\def\kpn{K^+\rightarrow\pi^+\nu\bar\nu}
\def\klpn{K_{L}\rightarrow\pi^0\nu\bar\nu}
\def\ksm{K_S\to\mu^+\mu^-}

\usepackage{fancyhdr}
\advance \headheight by 3.0truept       
\begin{document}

\begin{flushright}
    AJB-23-1\\
    {BARI-TH/23-743}
\end{flushright}

\medskip

\begin{center}
{\Large\bf
  \boldmath{331 Model Predictions for Rare $B$ and $K$ Decays,\\
    and $\Delta F=2$ Processes: an Update}}
\\[0.8 cm]
{\large\bf Andrzej~J.~Buras$^{a,b}$ and Fulvia~De~Fazio$^{c}$ 
 \\[0.5 cm]}
{\small
$^a$TUM Institute for Advanced Study, Lichtenbergstr. 2a, D-85747 Garching, Germany\\
$^b$Physik Department, Technische Universit\"at M\"unchen,
James-Franck-Stra{\ss}e, \\D-85747 Garching, Germany\\
$^c$Istituto Nazionale di Fisica Nucleare, Sezione di Bari, Via Orabona 4,
I-70126 Bari, Italy}
\end{center}

\vskip0.41cm


\begin{abstract}
\noindent
Motivated by the improved results from the HPQCD lattice collaboration
  on the hadronic matrix elements entering  $\Delta M_{s,d}$ in
  $B_{s,d}^0-\bar B_{s,d}^0$ mixings and the increase of the experimental
  branching ratio for $B_s\to\mu^+\mu^-$, we update our 2016 analysis of various flavour observables   in four  331 models, M1, M3, M13 and M16 based on the gauge group $SU(3)_C\times SU(3)_L\times U(1)_X$. These four models, which are distinguished by the quantum numbers, are selected among 24 331 models through their
  consistency with the electroweak precision tests and simultaneously
  by the relation  $C_9^\text{NP}=-b\, C_{10}^\text{NP}$ with $2\le b\le 5$,
which after
  new result on  $B_s\to\mu^+\mu^-$ from CMS is favoured over the
  popular relation $C_9^\text{NP}=- C_{10}^\text{NP}$  predicted by several
  leptoquark models.  In this context we investigate in particular
    the dependence of various observables on $\vcb$, varying  it in the broad range $[0.0386,\,0.043]$, that encompasses  both its inclusive and exclusive determinations. Imposing the experimental constraints from $\varepsilon_K$, $\Delta M_s$, $\Delta M_d$ and the mixing induced CP asymmetries $S_{\psi K_S}$ and $S_{\psi K_S}$, we investigate for which values of $\vcb$
    the four models can be made compatible with these data and what is the
    impact on $B$ and $K$ branching ratios. In particular we analyse
    NP contributions to the Wilson coefficients $C_9$ and $C_{10}$ and
    the decays $B_{s,d}\to\mu^+\mu^-$, $\kpn$ and $\klpn$.
  This allows us
  to illustrate how the value of $\vcb$ determined together with other
  parameters of these models is infected  by NP contributions and compare it
  with the one obtained recently under the assumption of the absence of NP
  in $\varepsilon_K$, $\Delta M_s$, $\Delta M_d$ and $ S_{\psi K_S}$.
  
\end{abstract}

\thispagestyle{empty}
\newpage
\setcounter{page}{1}

\tableofcontents

\section{Introduction}

The Standard Model (SM) describes globally the existing data on quark-flavour violating processes rather well \cite{Buras:2020xsm} but with the reduction 
of experimental errors and increased precision in non-perturbative and 
perturbative QCD and electroweak calculations a number of tensions
 at the level of $2-5\, \sigma$ seem to emerge in various seemingly unrelated 
observables. While some of these tensions could turn out to be the result of statistical 
fluctuations, underestimate of systematical and theoretical errors, it is not
excluded that eventually they all signal the presence of some kind of new physics (NP). Therefore, it is interesting to investigate what this NP could be.

In the present paper we will address some of these tensions in four particular 
331 models based on the gauge group $SU(3)_C\times SU(3)_L\times U(1)_X$   \cite{Pisano:1991ee,Frampton:1992wt} \footnote{A recent critical reanalysis of 331 models and a collection of references can be found in \cite{Pleitez:2021abk}. For a recent analysis see also \cite{CarcamoHernandez:2022fvl}.}. As these models have  much smaller number of 
new parameters than supersymmetric models, Randall-Sundrum scenarios and 
Littlest Higgs models, it is not evident that they can remove all present 
tensions simultaneously. 

Our paper has been motivated by the following recent facts.
\begin{itemize}
  \item
As demonstrated  in \cite{Buras:2022wpw}
most recent lattice QCD results from HPQCD collaboration 
\cite{Dowdall:2019bea}, based on $2+1+1$ simulations, imply
simultaneous agreement of
\be\label{loop}
|\varepsilon_K|,\qquad \Delta M_s,\qquad \Delta M_d, \qquad S_{\psi K_S}\,\qquad
S_{\psi \phi}
\ee
within the SM with the data for rather precise values of $\vcb$, $\vub$ and $\gamma$.
This should be contrasted with the situation at the time of our previous
analysis 2016 \cite{Buras:2016dxz}, when significant tensions between  
$\varepsilon_K$ and $\Delta M_{s,d}$ within the SM have been found
\cite{Blanke:2016bhf} and the room for NP in the quark mixing sector was much
larger than it is now.
\item
  The most recent data on $B_s\to\mu^+\mu^-$ from CMS imply that in the case of the dominance of left-handed quark currents, as is the case of the 331 models,
roughly \cite{Gubernari:2022hxn}
\be\label{relation}
C_9^\text{NP}=-b\, C_{10}^\text{NP},\qquad   {2\le b\le 5},
\ee
 where $C_9^\text{NP},\, C_{10}^\text{NP}$ represent the shifts in the Wilson coefficients $C_9,\,C_{10}$ of the $b \to s \ell^+ \ell^-$ effective Hamiltonian  in the presence of NP.
The relation (\ref{relation}) is in contrast to the previously favoured  case $b=1$ found in  several leptoquark models, in particular in the $U_1$ model. 
\item
 Recent messages from the LHCb  \cite{LHCb:2022qnv,LHCb:2022zom}, that the lepton flavour universality
   violation (LFUV) in $b\to s\ell^+\ell^-$, which for many years dominated the $B$-physics anomalies, practically disappeared. This is good news for 331 models for which LFUV  anomalies were problematic, although these models could provide some
   shifts in the Wilson coefficients $C_9$ and $C_{10}$. {Such shifts, in particular in $C_9$, are still required to describe suppressed branching ratios in
     $b\to s\mu^+\mu^-$ transitions.}
\item
  The most recent value for $\gamma$ obtained by the LHCb collaboration from tree-level decays that reads
 \cite{LHCb:2021dcr} 
\be\label{gamma}
  \gamma = (63.8^{+3.5}_{-3.7})^\circ \,.
  \ee
  It is significantly more precise than the LHCb values of $\gamma$ in 2016 that could be
  as large as $75^\circ$.
\end{itemize}

 The question then arises how  
 331 models face this new situation relative to the 2016 input and what are the implications for many flavour observables, in particular for the decays
 $B_d\to K(K^*)\mu^+\mu^-$, {$B^+\to K^+\mu^+\mu^-$ and $B_s\to \phi\mu^+\mu^-$ } related to the $B$ physics anomalies that imply
 the need for  significant NP contributions to the Wilson coefficient $C_9$ and smaller to $C_{10}$. 
 But it is also of interest to see what are
 the implications for rare decays $B_{s,d}\to\mu^+\mu^-$, $\kpn$ and $\klpn$.

It is known from many analyses, and stressed recently in particular in \cite{Buras:2021nns,Buras:2022wpw} that the tensions between inclusive and exclusive
determinations of $\vcb$ and $\vub$ preclude precise predictions for rare
decay observables in the SM. However, eliminating these parameters with the help of $\varepsilon_K$,  $\Delta M_{s,d}$ and $S_{\psi K_S}$ and setting the
latter observables to their experimental values allowed to obtain SM
predictions for many flavour observables that are most precise to date
 \cite{Buras:2021nns,Buras:2022wpw}.
 The motivation for this strategy has been strengthened recently by one of
  us \cite{Buras:2022qip} as the one which could minimize the impact of NP
  on the determination of the CKM parameters. Indeed, as demonstrated in 
\cite{Buras:2022wpw},   presently no NP
  is required to describe precise experimental data on $\Delta F=2$ observables.
  This allows in turn to determine the CKM parameters on the basis of $\Delta F=2$   observables alone without being involved in the issue of $\vcb$ and $\vub$ tensions and minimizing   possible impact of NP on their values that otherwise   would infect  SM predictions for rare decay branching ratios.

  The resulting values of the CKM parameters read
\cite{Buras:2022wpw}
\be\label{CKMoutput}
\boxed{\vcb=42.6(4)\times 10^{-3}, \qquad \vub=3.72(11)\times 10^{-3}, \qquad
\gamma=64.6(16)^\circ.}
\ee
While in this manner one can obtain rather precise SM predictions for numerous
branching ratios  \cite{Buras:2021nns,Buras:2022wpw,Buras:2022qip}, the absence
of NP in the $\Delta F=2$ observables, if confirmed with higher precision, would be  a {\em nightmare scenario}
for many NP models that attempt to explain the $B$ physics anomalies. While
the ones related to lepton flavour universality violation have been dwarfed recently through new LHCb data \cite{LHCb:2022qnv,LHCb:2022zom}, sizeable anomalies remained in several  branching ratios. In particular using the strategy of  \cite{Buras:2021nns,Buras:2022wpw}
large  anomalies in the low $q^2$ bin in
$B^+\to K^+\mu^+\mu^-$ ($5.1\sigma$) and $B_s\to \phi\mu^+\mu^-$ ($4.8\sigma$)
have been found \cite{Buras:2022qip}.

Explaining such anomalies without practically no NP contributions to $\Delta F=2$ processes is in principle possible but would require significant  tuning of NP
parameters. Now, the value of $\gamma$ in (\ref{CKMoutput}) agrees very well
with the most recent value from LHCb in (\ref{gamma}) and experimental value of
$\beta$ from  $S_{\psi K_S}$ is already used in obtaining the CKM parameters in
(\ref{CKMoutput}). It is evident then that the most efficient and transparent strategy to allow NP to enter the $\Delta F=2$ sector is to modify the value of $\vcb$.

In this context in \cite{Buras:2022wpw}, two scenarios for the
  parameters $\vcb$ and $\vub$ have been analysed within the SM. The EXCLUSIVE one based on determinations of these parameters in exclusive decays
\be\label{FLAGVUB1}
\vcb=39.21(62)\times 10^{-3},\qquad 
\vub=3.61(13)\times 10^{-3}, \qquad {(\rm EXCLUSIVE)},
\ee
and the HYBRID scenario in which the value for $\vcb$ is the inclusive one from \cite{Bordone:2021oof} and the exclusive
one for $\vub$ as above:
\be\label{HYBRID}
{\vcb=42.16(50)\times 10^{-3},\qquad 
\vub=3.61(13)\times 10^{-3}, \qquad {(\rm HYBRID)}.}
\ee

\begin{table}
\centering
\renewcommand{\arraystretch}{1.4}
\resizebox{\columnwidth}{!}{
\begin{tabular}{|l|lll|}
\hline
Decay 
& EXCLUSIVE
& HYBRID
&  DATA
\\
\hline \hline
 $\mathcal{B}(\kpn)\times 10^{11}$ & $6.88(38)$ & $8.44(41)$ & $10.9(38)$\hfill\cite{NA62:2022hqi} 
\\
 $\mathcal{B}(\klpn)\times 10^{11}$ & $2.37(15)$ & $2.74(14)$ & $< 300$ \hfill\cite{Ahn:2018mvc} 
\\
$\mathcal{B}(\ksm)\times 10^{13}_\text{SD}$ & $1.49(10) $ & $1.72(8)$ &
$< 0.8~10^4$\hfill\cite{Aaij:2017tia} 
\\
$\overline{\mathcal{B}}(B_s\to\mu^+\mu^-)\times 10^{9}$ & $3.18(12)$ &  $3.67(12)$
&   $3.45(29)$\hfill\cite{LHCb:2021awg,CMS:2020rox,ATLAS:2020acx,HFLAV:2022pwe}
\\
$\mathcal{B}(B_d\to\mu^+\mu^-)\times 10^{10}$ & $0.864(34) $ & $0.999(34)$
&  $<2.05$\hfill\cite{LHCb:2021awg}
\\
$ |\varepsilon_K|\times 10^{3}$ & $1.78(11)$ &  $2.14(12)$  & $ 2.228(11)$\hfill\cite{Zyla:2020zbs}
\\
$S_{\psi K_S}$  & $0.731(24)$ &  $0.688(22)$  & $0.699(17)$\hfill\cite{Zyla:2020zbs}
\\
 $\Delta M_s \,\text{ps}^{-1}$ &$15.02(87)$ &  $17.35(94)$   & $17.749(20)$\hfill \cite{Zyla:2020zbs}
\\
 $\Delta M_d \,\text{ps}^{-1}$ &$0.434(28)$ &  $ 0.502(31) $  & $ 0.5065(19)$\hfill \cite{Zyla:2020zbs}
\\
\hline
\end{tabular}
}
\renewcommand{\arraystretch}{1.0}
\caption{\label{tab:SMBRs}
  \small
  Predictions (second column) for selected observables within the SM obtained in  \cite{Buras:2022wpw} using the EXCLUSIVE strategy for $\vcb$ and $\vub$ and $\gamma=65.4^\circ$. In the third column we show the results for the HYBRID choice of $\vcb$ and $\vub$ as given in (\ref{HYBRID}) and in the fourth the experimental data.
}
\end{table}

In Table~\ref{tab:SMBRs} we show selected results obtained in  \cite{Buras:2022wpw} in these two scenarios. The results obtained in the HYBRID scenario
  do not differ by much from those obtained using the CKM parameters in (\ref{CKMoutput}) \cite{Buras:2022wpw,Buras:2022qip}.
  With exclusive values of $\vcb$ that are much lower than given in
(\ref{CKMoutput}),  anomalies
in $\Delta M_s$ ($3\sigma$), $\Delta M_d$ ($4\sigma$) and $\varepsilon_K$  ($5\sigma$) are generated.
But in \cite{Buras:2022wpw} no analysis of a NP scenario has been presented
which would explain these anomalies and whether a model explaining them would also be able to explain anomalies in semi-leptonic B decays. In the present paper we investigate
whether the 331 models could provide some insight in these issues
 and what would
 be the implications for rare branching ratios. Our  analysis
 illustrates in simple settings how the determination of $\vcb$ in a global fit
 that includes observables exposing anomalies can be infected by NP contributions \cite{Buras:2022qip}. Indeed the allowed values of $\vcb$ depend on the 331
 model considered. It is a concrete illustration of the points made
   in section 2 of the latter paper.

Our paper is organized as follows.
In Section~\ref{sec:2} we recall briefly the flavour structure of the
331 models. 
In Section~\ref{Selection} we select
four  331 models that perform best on the basis of electroweak precision tests
and the present experimental values of the ratio $C_9^\text{NP}/C_{10}^\text{NP}$
in (\ref{relation}). In fact these are the only  models among the 24
ones considered in  \cite{Buras:2014yna}, that 
can successfully face the new relation (\ref{relation})
when other constraints like electroweak precision tests are taken into account
\cite{Buras:2016dxz}.
In Section~\ref{sec:3} we present numerical analysis of these models {addressing the issues mentioned above.}
We conclude in Section~\ref{sec:5}.

\section{Flavour Structure of 331 Models}\label{sec:2}
Let us recall that in the 331 models new flavour-violating effects are 
governed by tree-level $Z^\prime$ exchanges with a subdominant but non-negligible  role played 
by tree-level $Z$ exchanges generated through $Z-Z^\prime$ mixing. All the 
formulae for flavour observables in these models can be found in 
 \cite{Buras:2012dp,Buras:2013dea,Buras:2014yna,Buras:2015kwd} and will not 
be repeated here. In particular the collection of formulae for $Z^\prime$ couplings to quarks and leptons 
are given in \cite{Buras:2013dea}.

New sources of flavour and CP violation in 331 models are parametrized by
new mixing parameters and phases
\be\label{PAR}
\tilde s_{13},\qquad\tilde s_{23},\qquad  \delta_1,\qquad \delta_2
\ee
with $\tilde s_{13}$ and $\tilde s_{23}$ positive definite  and smaller than  unity and 
$0\le \delta_{1,2}\le 2\pi$. They can be constrained by flavour observables as demonstrated in detail in \cite{Buras:2012dp}. 
The non-diagonal $Z^\prime$ couplings  relevant for $K$, $B_d$ and $B_s$ meson 
systems can 
be then parametrized respectively within an excellent approximation through
\be\label{vij}
 v_{32}^*v_{31}=\tilde s_{13}\tilde s_{23}e^{i(\delta_2-\delta_1)}, \qquad
 v_{33}^*v_{31}=-\tilde s_{13}e^{-i\delta_1}, \qquad
 v_{33}^*v_{32}=-\tilde s_{23}e^{-i\delta_2} \,.
\ee
$\tilde s_{13}$ and $\delta_1$ can be determined from $\Delta M_d$ and CP-asymmetry $S_{\psi K_S}$ while $\tilde s_{23}$ and $\delta_2$ from $\Delta M_s$ and CP-asymmetry $S_{\psi \phi}$. Then the parameters in the $K$ system are fixed.
 It is a remarkable feature of 331 models that also FCNC processes  in the charm sector can be described without introducing no new free parameters beyond those already
present in the beauty  and kaon meson systems \cite{Colangelo:2021myn,Buras:2021rdg}.  These correlations constitute important tests of these models.

The remaining two parameters, except for $M_{Z^\prime}$ mass, are
$\beta$ and $\tan\bar\beta$ defined through\footnote{The parameter $\beta$ should not be confused with the angle $\beta$ in the unitarity triangle.}
\be\label{QTX}
Q=T_3+\frac{Y}{2}= T_3+\beta T_8+X, \qquad \tan\bar\beta=\frac{v_\rho}{v_\eta}\,.
\ee
Here $T_{3,8}$ and $X$ are the  diagonal generators of  $SU(3)_L$ and 
$U(1)_X$, respectively. $Y$ represents $U(1)_Y$ and $v_i$ are  
the vacuum expectation values of scalar triplets responsible for the generation of down- and up-quark masses in these models.

Different 331 models can also be distinguished by the way quarks transform
under  $SU(3)_L$. In \cite{Buras:2014yna} two classes of such models
have been analyzed to be denoted by $F_1$ and $F_2$.
$F_1$ stands for the case in which  
the first two generations of quarks belong to triplets of $SU(3)_L$,
while the third generation of quarks to antitriplet. $F_2$
stands for the case in which  
the first two generations of quarks belong to antitriplets of $SU(3)_L$,
while the third generation of quarks to triplet.

A detailed analysis of 24 331 models corresponding to  different values of 
$\beta$ and $\tan\bar\beta$ for the representations $F_1$ and $F_2$ has been presented in \cite{Buras:2014yna}. They are collected in Table~\ref{tab:331models}. With the values of  $\beta$ and $\tan\bar\beta$ being fixed, flavour phenomenology depends only on the parameters in (\ref{PAR}),  $M_{Z^\prime}$ and the CKM parameters  which distinguish EXCLUSIVE and HYBRID scenarios.

\section{Selecting the 331 Models}\label{Selection}
A detailed analysis of electroweak precision tests in the 24
models  in Table~\ref{tab:331models}   has been performed in  \cite{Buras:2014yna}. Interested readers are asked to look at Section~5 of that paper. Here we just summarize the main outcome of that study.

Requiring that the 24 models in question perform well in
these tests and are simultaneously consistent with the ratio $C_9/C_{10}$ in
(\ref{relation}) selects, as shown in Table~\ref{tab:C9C10},
the following models
\be\label{favoured}
{\rm M1}, \qquad {\rm M3},\qquad {\rm M13}, \qquad {\rm M16}, \qquad {(\rm favoured)}.
\ee
Note that the  $Z-Z^\prime$ mixing
plays in some cases an important role and that the two favoured models M8 and M9
analysed by us in \cite{Buras:2016dxz} are ruled out by (\ref{relation}).

\begin{table}[!tb]
{\renewcommand{\arraystretch}{1.3}
\begin{center}
\begin{tabular}{|c||c|c|c||c||c|c|c||c||c|c|c|}
\hline
MI  &     {\rm scen.} &  $\beta$ & $\tan {\bar \beta}$ & MI  & {\rm scen.} &  $\beta$ & $\tan {\bar \beta}$ &MI  & {\rm scen.} &  $\beta$ & $\tan {\bar \beta}$\\
\hline
M1 & $F_1$ & $-2/\sqrt{3}$ & 1 & M9 & $F_2$ & $-2/\sqrt{3}$ & 1 & M17 & $F_1$ & $-2/\sqrt{3}$ & 0.2
\\
M2 & $F_1$ & $-2/\sqrt{3}$ & 5 & M10 & $F_2$ & $-2/\sqrt{3}$ & 5 & M18 & $F_2$ & $-2/\sqrt{3}$ & 0.2
\\
M3 & $F_1$ & $-1/\sqrt{3}$ & 1 & M11 & $F_2$ & $-1/\sqrt{3}$ & 1 & M19 & $F_1$ & $-1/\sqrt{3}$ & 0.2
\\
M4 & $F_1$ & $-1/\sqrt{3}$ & 5 & M12 & $F_2$ & $-1/\sqrt{3}$ & 5 & M20 & $F_2$ & $-1/\sqrt{3}$ & 0.2
\\
M5 & $F_1$ & $1/\sqrt{3}$ & 1 & M13 & $F_2$ & $1/\sqrt{3}$ & 1 & M21 & $F_1$ & $1/\sqrt{3}$ & 0.2
\\
M6 & $F_1$ & $1/\sqrt{3}$ & 5 & M14 & $F_2$ & $1/\sqrt{3}$ & 5 & M22 & $F_2$ & $1/\sqrt{3}$ & 0.2
\\
M7 & $F_1$ & $2/\sqrt{3}$ & 1 & M15 & $F_2$ & $2/\sqrt{3}$ & 1 & M23 & $F_1$ & $2/\sqrt{3}$ & 0.2
\\
M8 & $F_1$ & $2/\sqrt{3}$ & 5 & M16 & $F_2$ & $2/\sqrt{3}$ & 5 & M24 & $F_2$ & $2/\sqrt{3}$ & 0.2
\\
\hline
\end{tabular}
\end{center}}
\caption{\it Definition of the various 331 models.
\label{tab:331models}}~\\[-2mm]\hrule
\end{table}

\begin{table}[!tb]
{\renewcommand{\arraystretch}{1.3}
\begin{center}
\begin{tabular}{|c||c|c||c||c|c||c||c|c|}
\hline
MI  &     {\rm Full} &  {\rm no Mixing} & MI  & {\rm Full} &  {\rm no Mixing}
& MI  & {\rm Full} &  {\rm no Mixing}\\
\hline
M1 & $-3.25$ & $-8.87$ & M9 & $0.42$ &$0.60$ & M17 & $-175.6$ & $-8.87$
\\
M2 & $-1.68$ & $-8.87$ & M10 & $0.28$ &$0.60$ & M18 & $0.75$ & $0.60$
\\
M3 & $-2.07$ & $-2.98$ & M11 & $-0.02$ &$-0.004$ & M19 & $-63.48$ & $-2.98$
\\
M4 & $-1.09$ & $-2.98$ & M12 & $-0.04$ &$-0.004$ & M20 & $0.06$ & $-0.004$
\\
M5 & $0.02$ & $-0.004$ & M13 & $-5.47$ & $-2.98$ & M21 & $1.15$ & $-0.004$
\\
M6 & $-0.03$ & $-0.004$ & M14 & $-1.56$ & $-2.98$ & M22 & $3.25$ & $-2.98$
\\
M7 & $0.97$ & $0.60$ & M15 & $11.3$ & $-8.87$ & M23 & $7.50$ & $0.60$
\\
M8 & $0.49$ & $0.60$ & M16 & $-4.59$ & $-8.87$ & M24 & $2.44$ & $-8.87$\\
\hline
\end{tabular}
\end{center}}
\caption{\it $C_9^\text{NP}/C_{10}^\text{NP}$ in  various 331 models with and without $Z-Z^\prime$ mixing for $M_{Z^\prime}=3\tev$.
\label{tab:C9C10}}~\\[-2mm]\hrule
\end{table}

\section{Numerical Analysis  }\label{sec:3}

\begin{table}[!tb]
\center{\begin{tabular}{|l|l|}
\hline
$m_{B_s} = 5366.8(2)\mev$\hfill\cite{Zyla:2020zbs}	&  $m_{B_d}=5279.58(17)\mev$\hfill\cite{Zyla:2020zbs}\\
$\Delta M_s = 17.749(20) \,\text{ps}^{-1}$\hfill \cite{Zyla:2020zbs}	&  $\Delta M_d = 0.5065(19) \,\text{ps}^{-1}$\hfill \cite{Zyla:2020zbs}\\
{$\Delta M_K = 0.005292(9) \,\text{ps}^{-1}$}\hfill \cite{Zyla:2020zbs}	&  {$m_{K^0}=497.61(1)\mev$}\hfill \cite{Zyla:2020zbs}\\
$S_{\psi K_S}= 0.699(17)$\hfill\cite{Zyla:2020zbs}
		&  {$F_K=155.7(3)\mev$\hfill  \cite{Aoki:2019cca}}\\
	$|V_{us}|=0.2253(8)$\hfill\cite{Zyla:2020zbs} &
 $|\eps_K|= 2.228(11)\cdot 10^{-3}$\hfill\cite{Zyla:2020zbs}\\
$F_{B_s}$ = $230.3(1.3)\mev$ \hfill \cite{Aoki:2021kgd} & $F_{B_d}$ = $190.0(1.3)\mev$ \hfill \cite{Aoki:2021kgd}  \\
$F_{B_s} \sqrt{\hat B_s}=256.1(5.7) \mev$\hfill  \cite{Dowdall:2019bea}&
$F_{B_d} \sqrt{\hat B_d}=210.6(5.5) \mev$\hfill  \cite{Dowdall:2019bea}
\\
 $\hat B_s=1.232(53)$\hfill\cite{Dowdall:2019bea}        &
 $\hat B_d=1.222(61)$ \hfill\cite{Dowdall:2019bea}          
\\
{$m_t(m_t)=162.83(67)\gev$\hfill\cite{Brod:2021hsj} }  & {$m_c(m_c)=1.279(13)\gev$} \\
{$S_{tt}(x_t)=2.303$} & {$S_{ut}(x_c,x_t)=-1.983\times 10^{-3}$} \\
    $\eta_{tt}=0.55(2)$\hfill\cite{Brod:2019rzc} & $\eta_{ut}= 0.402(5)$\hfill\cite{Brod:2019rzc}\\
$\kappa_\varepsilon = 0.94(2)$\hfill \cite{Buras:2010pza}	&
$\eta_B=0.55(1)$\hfill\cite{Buras:1990fn,Urban:1997gw}\\
$\tau_{B_s}= 1.515(4)\,\text{ps}$\hfill\cite{Amhis:2016xyh} & $\tau_{B_d}= 1.519(4)\,\text{ps}$\hfill\cite{Amhis:2016xyh}   
\\	       
\hline
\end{tabular}  }
\caption {\textit{Values of the experimental and theoretical
    quantities used as input parameters. For future 
updates see FLAG  \cite{Aoki:2021kgd}, PDG \cite{Zyla:2020zbs}  and HFLAV  \cite{Aoki:2019cca}. 
}}
\label{tab:input}
\end{table}

\subsection{Determining the parameter space}

Despite the fact that NP is not required to obtain within the SM simultaneous agreement   with data for the $\Delta F=2$ observables in (\ref{loop}) \cite{Buras:2022wpw}, the present uncertainties in hadronic parameters still allow
  for some NP contributions, whose size depends strongly on the value of $\vcb$
  \cite{Buras:2021nns,Buras:2022wpw}. Therefore   in order to constrain the parameters in (\ref{PAR}) and subsequently obtain predictions for various observables, we will proceed in each of the four considered 331 models
  as follows:
  \begin{itemize}
  \item
    We will vary $\Delta M_d,\,S_{\psi K_s},\, \Delta M_s,\,S_{\psi \phi},\,\epsilon_K$ within $5\%$ of the central value of their experimental datum. This
      amount is based on the uncertainties in the CKM parameters given in (\ref{CKMoutput}) determined  using SM expressions for the observables in question.
      They are generally below $5\%$, typically $(2-3)\%$ but as they follow
      dominantly from uncertainties of hadronic matrix elements, which could
      still be modified, we use $5\%$ to be conservative.
        \item
    Concerning CKM parameters, we adopt here a different strategy with respect to our previous analyses. We vary  $|V_{ub}|$ as in (\ref{CKMoutput}), while  $|V_{cb}|$ is varied in such a way to encompass  both its inclusive and exclusive determinations, i.e. $|V_{cb}| \in [ 0.0386,\,0.043]$. 
  \item
    For each of the four 331 models considered in this paper we then  determine the allowed values of the 331 parameters $\tilde s_{13},\, \delta_1,\,\tilde s_{23},\,\delta_2$ as well as a range for $|V_{cb}|$ for which a given model satisfies
    the constraints from $\Delta F=2$ observables in (\ref{loop}) within
    $5\%$ as stated above.
  \item
    We  predict   several observables in each model and discuss their dependence on  $|V_{cb}|$. We compare the outcome in the four cases.
  \end{itemize}
  
The remaining parameters used in our analysis  are collected in Table~\ref{tab:input}.

\begin{figure}[!tb]
\begin{center}
\includegraphics[width = 0.8\textwidth]{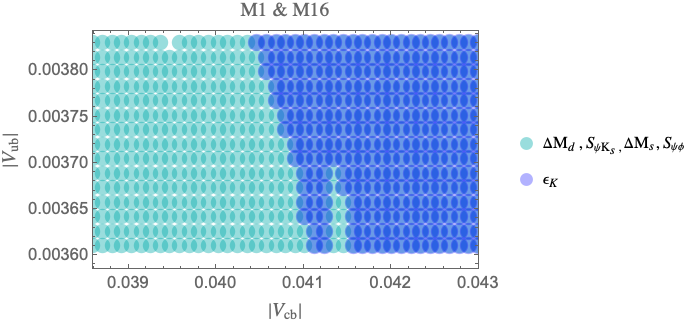}\\
\vskip 0.5cm
\includegraphics[width = 0.8\textwidth]{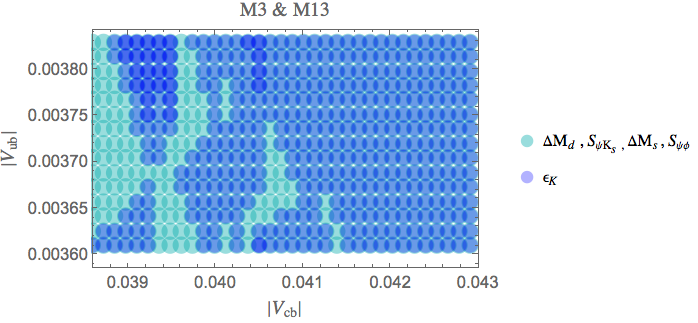}
    \caption{\small Allowed $(|V_{cb}|,|V_{ub}|)$ ranges in the  parameter space of M1 and M16 (upper plot) and in that of M3 and M13 (lower plot). Each point corresponds to a set of 331 parameters. The green points are obtained after imposing the constraints on  $\Delta M_d,\,S_{\psi K_s},\, \Delta M_s,\,S_{\psi \phi}$, while the light blue points derive from imposing the constraint on $\varepsilon_K$. }\label{figCKM}
\end{center}
\end{figure}

Among the parameters that define the various scenarios, $\Delta F=2$ observables depend only on $|\beta|$, so that the resulting parameter space will be the same for M1 and M16 as well as for M3 and M13.
In the two cases we have constructed the tables of the allowed parameters in the form of 6-vectors of the kind $(\tilde s_{13},\, \delta_1,\,\tilde s_{23},\,\delta_2,\,|V_{cb}|,\,|V_{ub}|)$. Of course it is not possible to display the space of all the variables simultaneously and therefore we do not show these plots. Instead, in Fig. \ref{figCKM} we show the allowed $(|V_{cb}|,|V_{ub}|)$ ranges in the two resulting parameter spaces. It should be understood that each point corresponds to a set of 331 parameters. In these figures the green points are obtained after imposing the constraints on  $\Delta M_d,\,S_{\psi K_s},\, \Delta M_s,\,S_{\psi \phi}$ and show that even though such observables select the 331 parameters $\tilde s_{13},\, \delta_1,\,\tilde s_{23},\,\delta_2$ they do not have an impact on the allowed ranges for $|V_{ub}|$ and $|V_{cb}|$. On the contrary, when the constraint on $\varepsilon_K$ is imposed, a limitation is found for $\vcb$ that is the consequence of the stronger dependence of $\varepsilon_K$ on this parameter than
  in the case of $\Delta M_s$ and $\Delta M_d$.
However, we can observe that, while in the case of M1 and M16, $|V_{cb}|$ cannot be smaller than $\simeq 0.0405$,  no similar constraint is found in the case of M3, M13.

\subsection{ $C_9^\text{NP}$ and $C_{10}^\text{NP}$}
\begin{figure}[!tb]
\begin{center}
\includegraphics[width = 0.9\textwidth]{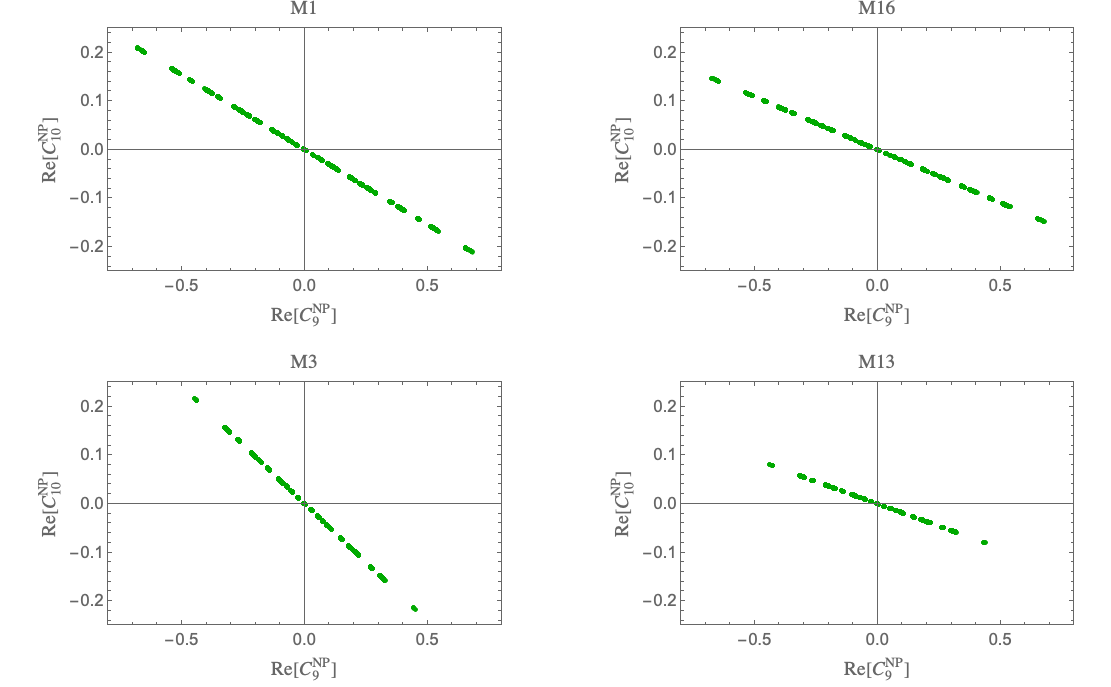}
    \caption{\small  Correlation between the real parts of $C_9^{NP}$ and $C_{10}^{NP}$ in the four considered 331 models.}\label{figC9C10}
\end{center}
\end{figure}

\begin{figure}[!tb]
\begin{center}
\includegraphics[width = 0.9\textwidth]{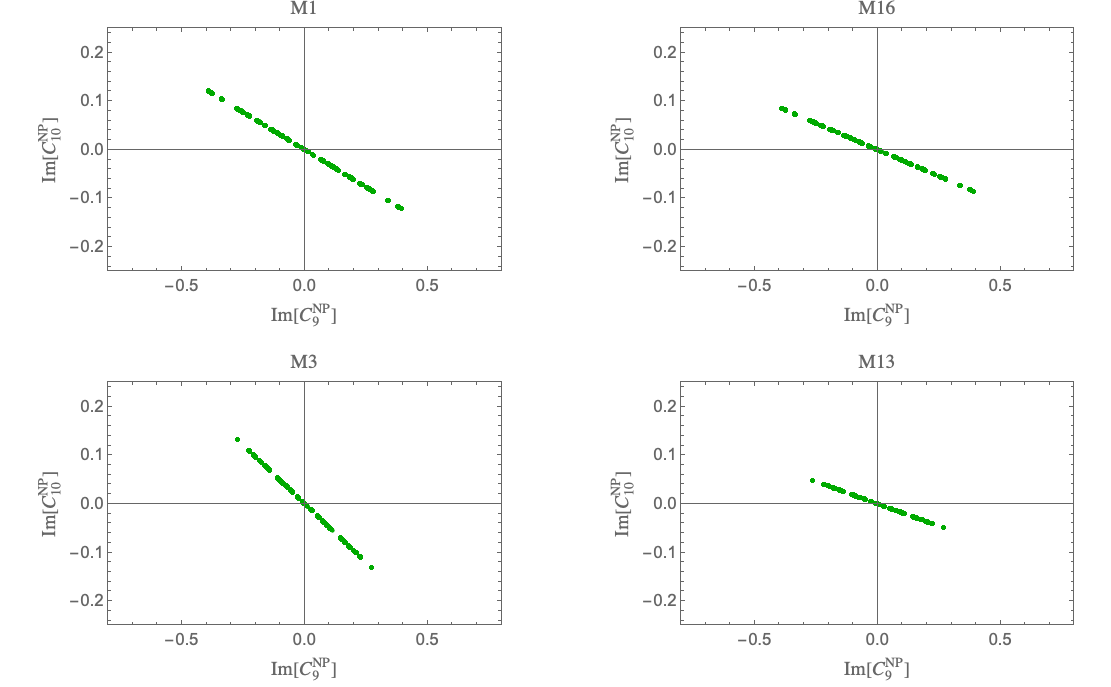}
    \caption{\small  Correlation between the imaginary parts of $C_9^{NP}$ and $C_{10}^{NP}$ in the four considered 331 models.}\label{figImC9C10}
\end{center}
\end{figure}

We have already remarked the nice feature of 331 models that the ratio $C_9^\text{NP}/C_{10}^\text{NP}$ depends only on the considered scenario but not on the parameters $\tilde s_{13},\, \delta_1,\,\tilde s_{23},\,\delta_2$. However, the separate values of $C_9^\text{NP}$ and $C_{10}^\text{NP}$ depend on them. In Fig. \ref{figC9C10} we show the correlation between their real parts in the four scenarios, while in Fig. \ref{figImC9C10} the  correlation between their imaginary parts is displayed.

\begin{figure}[!tb]
\begin{center}
\includegraphics[width = 0.9\textwidth]{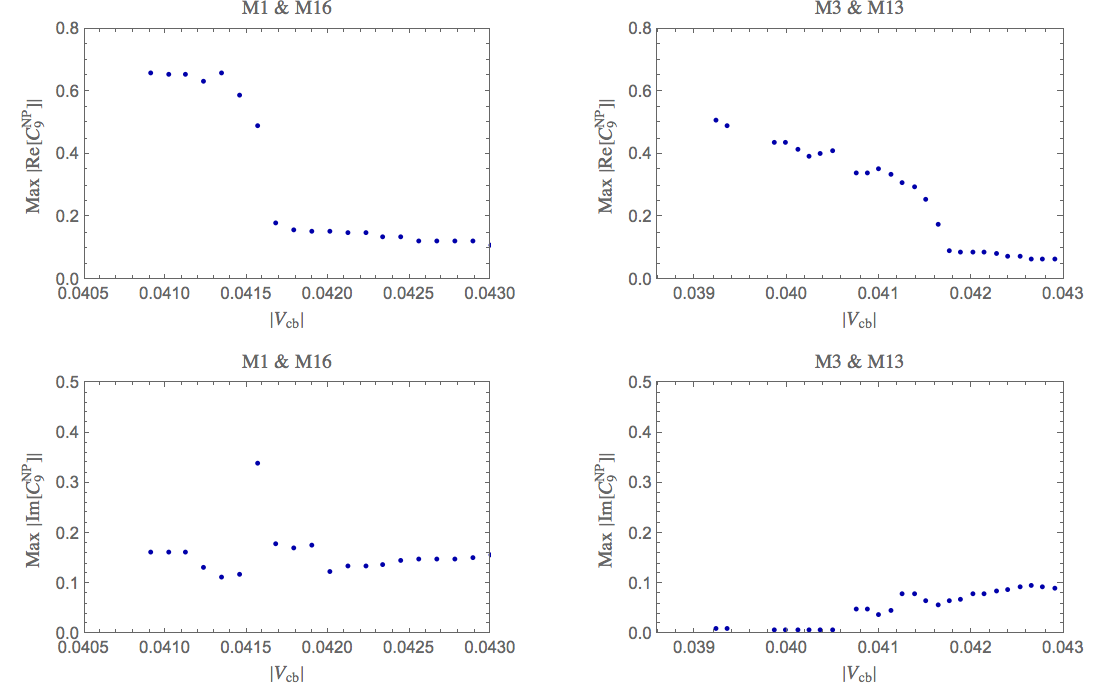}
    \caption{\small  Maximal deviation of $\left| {\rm Re} [C_9^{NP}] \right|$ and  $\left| {\rm Im} [C_9^{NP}] \right|$ in the four considered 331 models. }\label{figC9Vcb}
\end{center}
\end{figure}

In order to understand which values of $|V_{cb}|$ correspond to the largest deviations in $C_9^\text{NP}$ we consider ${\rm Max} \left|{\rm Re}[C_9^\text{NP}] \right|$ setting $|V_{ub}|$ at its central value. The result is shown in Fig. \ref{figC9Vcb}.
\begin{figure}[!tb]
\begin{center}
\includegraphics[width = 0.9\textwidth]{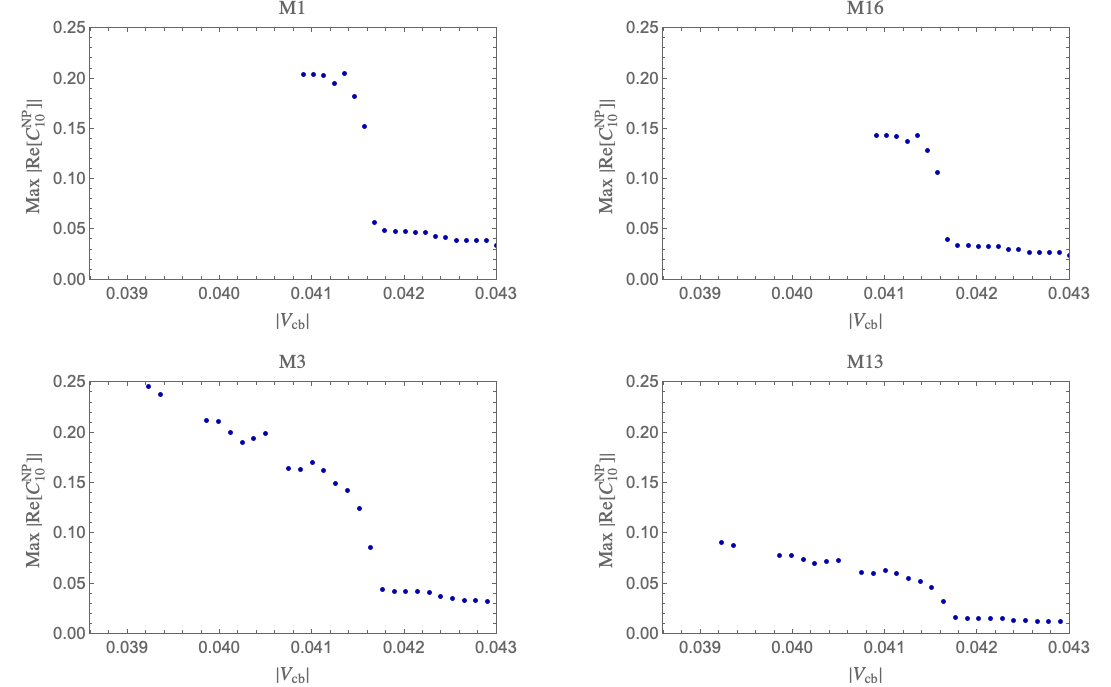}
    \caption{\small  Maximal deviation of $\left| {\rm Re} [C_{10}^{NP}] \right|$  in the four considered 331 models. }\label{figC10Vcb}
\end{center}
\end{figure}
\begin{figure}[!tb]
\begin{center}
\includegraphics[width = 0.9\textwidth]{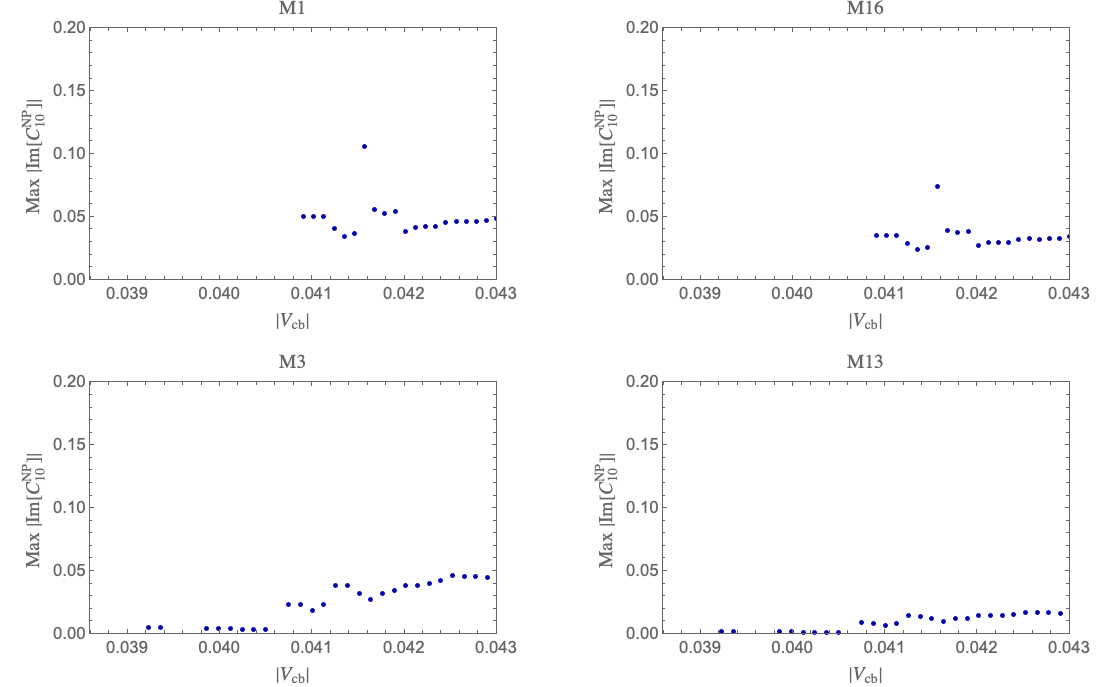}
    \caption{\small  Maximal deviation of $\left| {\rm Im} [C_{10}^{NP}] \right|$  in the four considered 331 models. }\label{figImC10Vcb}
\end{center}
\end{figure}
These plots  display  that, consistently with the result in Fig. \ref{figCKM}  in the case of M1 and M16 only the values  $|V_{cb}|\ge 0.0405$  are allowed. Moreover, the  deviation in $|{\rm Re}[C_9]|$ is a decreasing function of   $\vcb$, as shown in Fig. \ref{figC9Vcb}, together with the plots for the imaginary part.
This dependence on $\vcb$ follows from the fact, as seen in (\ref{CKMoutput}), that  the experimental value of $\Delta M_s$   is best reproduced within the SM for $|V_{cb}|\approx 0.0426$ so that the   room left for $Z^\prime$ contributions to $\Delta M_s$ decreases with   increasing $\vcb$ and in turn not allowing sizeable impact on $C_9$.

The  situation for $|{\rm Re}[C_{10}^{NP}]|$ and $|{\rm Im}[C_{10}^{NP}]|$ is displayed in Figs. \ref{figC10Vcb} and \ref{figImC10Vcb}.  It can be noticed that $C_9^{NP}$ is to an excellent approximation the same in M1 and M16 on the one hand and in M3 and M13 on the other; for this reason we have shown the corresponding plots in a single figure. $C_{10}^{NP}$ is instead different in all the four considered cases.

We observe that while the  pattern of NP contributions signalled by the data is correctly
  described by these models, the absolute values of $C_9^\text{NP}$ are likely
  to   turn out to be too small to explain the observed suppression of
  the branching ratios for
  $B^+\to K^+\mu^+\mu^-$  and $B_s\to \phi\mu^+\mu^-$, in particular if
  the final value for $\vcb$ from tree-level decays will turn out
  to be in the ballpark of its inclusive determinations.

\subsection{ ${\bar {\cal B}}(B_s \to \mu^+ \mu^-) $ and $ {\cal B}(B_d \to \mu^+ \mu^-) $}
\begin{figure}[!tb]
\begin{center}
\includegraphics[width = 0.39\textwidth]{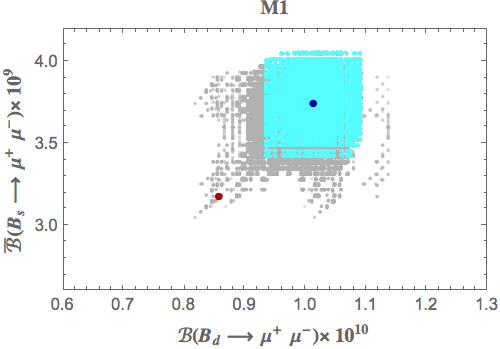}\hskip 0.5cm
\includegraphics[width = 0.54\textwidth]{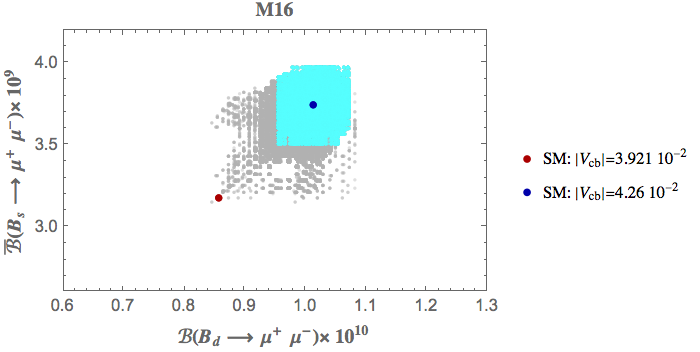}\\
\vskip 0.2cm
\includegraphics[width = 0.39\textwidth]{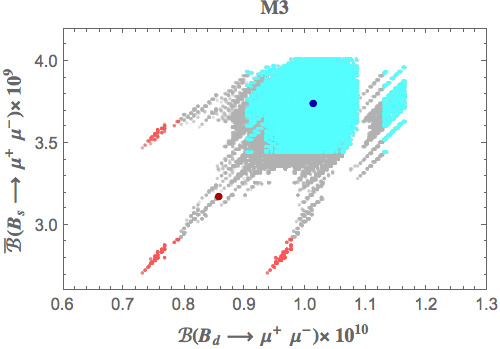}\hskip 0.5cm
\includegraphics[width = 0.54\textwidth]{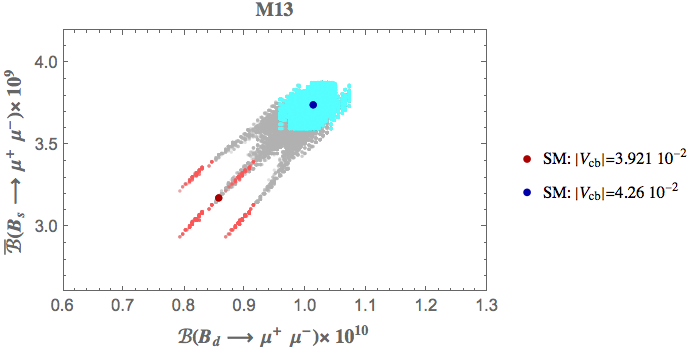}
    \caption{\small  Correlation between ${\bar {\cal B}}(B_s \to
     \mu^+ \mu^-) $ and $ {\cal B}(B_d \to \mu^+ \mu^-) $. The gray points span all the allowed parameter space in each scenario.
    The red region corresponds to $|V_{cb}| \in [0.0386,\,0.0398]$ while the cyan region corresponds to $|V_{cb}| \in [0.0422,\,0.043]$. The SM results in correspondence of two values of $|V_{cb}|$ are displayed, as specified in the legends.}\label{figBsvsBd}
\end{center}
\end{figure}

\begin{figure}[!tb]
\begin{center}
\includegraphics[width = 0.45\textwidth]{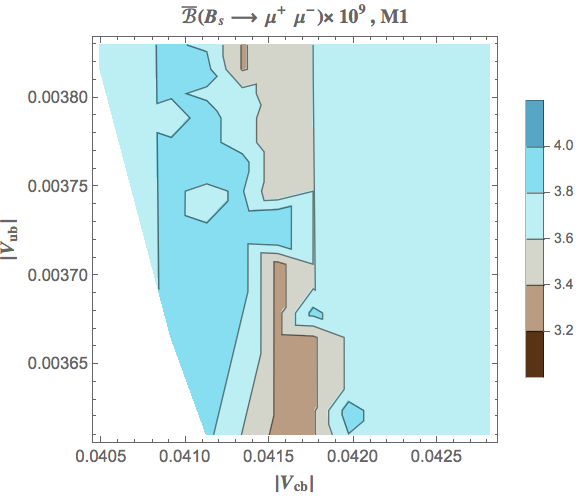}\hskip 0.5cm
\includegraphics[width = 0.45\textwidth]{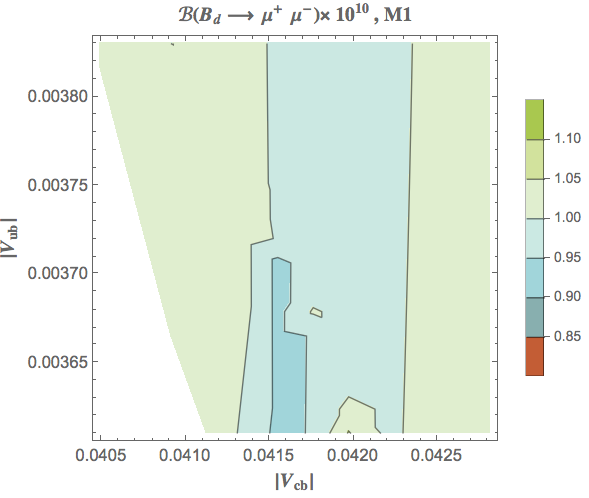}\\\vskip 0.2cm
\includegraphics[width = 0.45\textwidth]{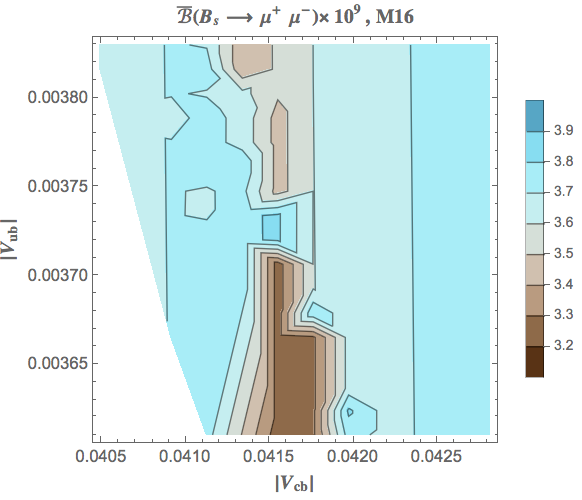}\hskip 0.5cm
\includegraphics[width = 0.45\textwidth]{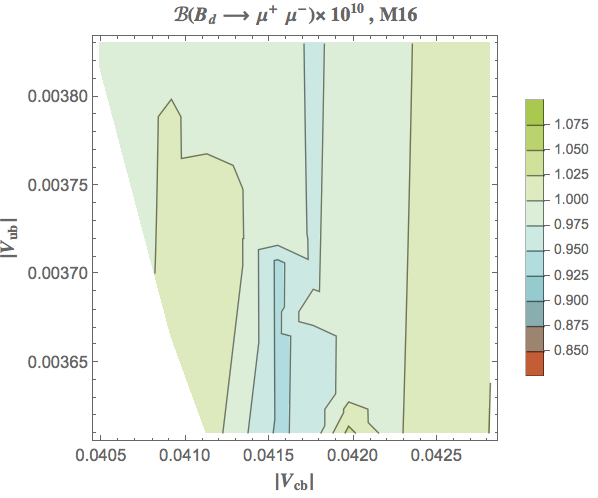}
    \caption{\small  Contour Plots of ${\bar {\cal B}}(B_s \to \mu^+ \mu^-) $ (left column) and $ {\cal B}(B_d \to \mu^+ \mu^-) $ (right column) versus $|V_{cb}|$ and $|V_{ub}|$ in M1 (upper plots) and in M16 (lower plots). }\label{figBs1CP}
\end{center}
\end{figure}

\begin{figure}[!tb]
\begin{center}
\includegraphics[width = 0.45\textwidth]{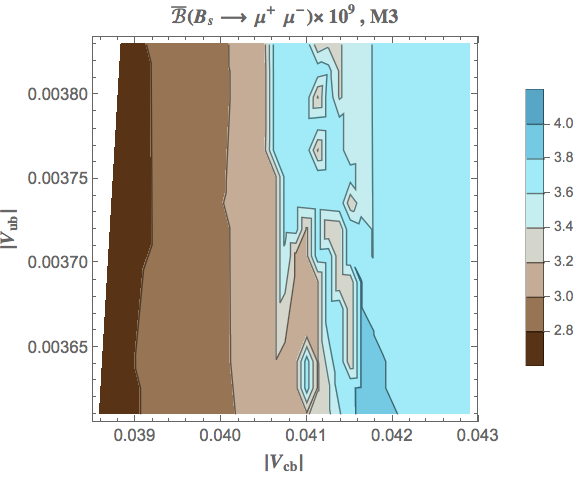}\hskip 0.5cm
\includegraphics[width = 0.45\textwidth]{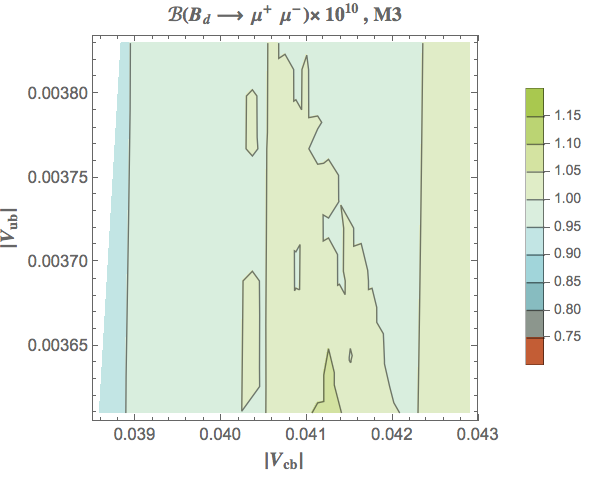}\\ \vskip 0.2cm
\includegraphics[width = 0.45\textwidth]{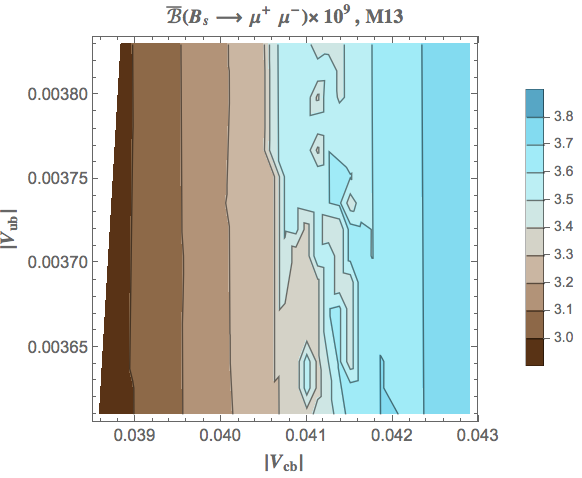}\hskip 0.5cm
\includegraphics[width = 0.45\textwidth]{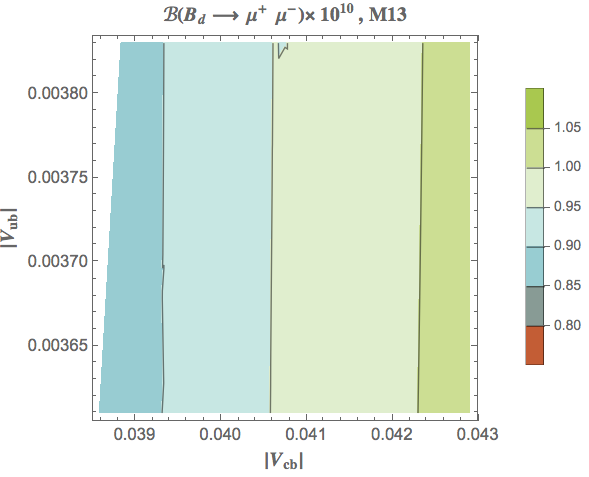}
    \caption{\small  Contour Plots of ${\bar {\cal B}}(B_s \to \mu^+ \mu^-) $ (left column) and $ {\cal B}(B_d \to \mu^+ \mu^-) $ (right column) versus $|V_{cb}|$ and $|V_{ub}|$ in M3 (upper plots)and  in M13  (lower plots). }\label{figBs3CP}
\end{center}
\end{figure}

\begin{figure}[!tb]
\begin{center}
\includegraphics[width = 0.45\textwidth]{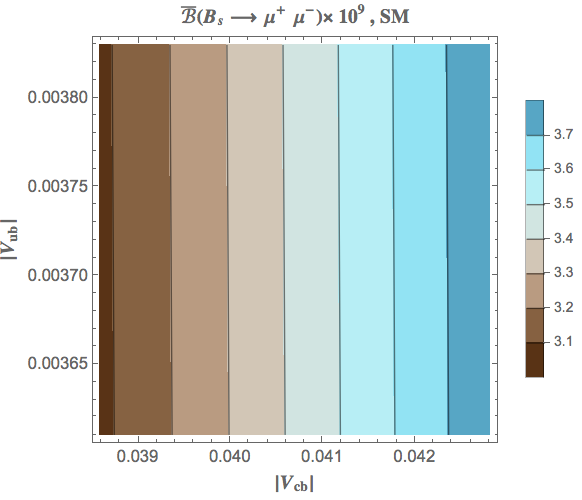}\hskip 0.5cm
\includegraphics[width = 0.45\textwidth]{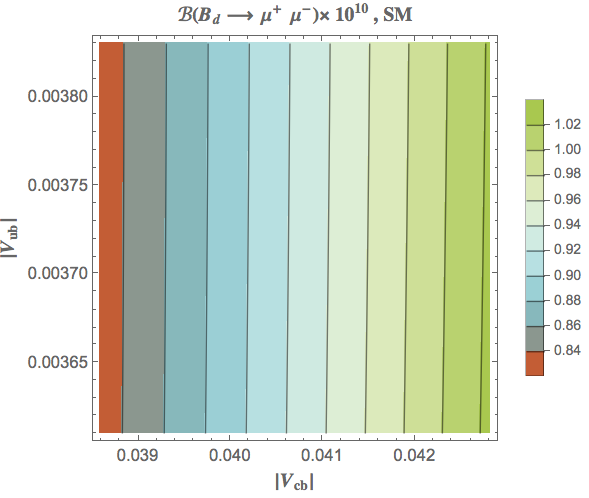} 
    \caption{\small  Contour Plots of ${\bar {\cal B}}(B_s \to \mu^+ \mu^-) $ (left column) and $ {\cal B}(B_d \to \mu^+ \mu^-) $ (right column) versus $|V_{cb}|$ and $|V_{ub}|$ in the  SM . }\label{figBsSMCP}
\end{center}
\end{figure}

In Fig. \ref{figBsvsBd} we plot the correlation between the rare decays ${\bar {\cal B}}(B_s \to \mu^+ \mu^-) $ and $ {\cal B}(B_d \to \mu^+ \mu^-) $ in the four considered 331 models. In these plots, the gray region is obtained considering all the allowed parameter space in each scenario, while the red region corresponds to $|V_{cb}| \in [0.0386,\,0.0398]$ and the cyan region  to $|V_{cb}| \in [0.0422,\,0.043]$. The SM results for {$\vcb=0.03921$ and $\vcb=0.0426$} are also displayed. Comparing the four models, we can observe that if $\vcb$ is fixed consistently with the exclusive determinations,
a possible suppression of both branching ratios with respect to their
SM values, {that is not yet excluded in view of large experimental errors,} could be explained only in M3 and M13. On the other hand, inclusive values of $\vcb$ do not define a clear situation in any of the four models: other correlations should be explored in order to discriminate among these scenarios.
We detail the dependence of the considered branching fractions on the CKM elements in the contour plots in Fig. \ref{figBs1CP} for M1 and M16 and in Fig. \ref{figBs3CP} for M3 and M13.  
Since in each scenario the parameter space involves 6 variables it is possible that fixing $(\vcb,\,\vub)$ different values for the considered branching ratios are obtained, because these depend also on the other four parameters of the 331 model. Therefore, what is plotted in Fig. \ref{figBs1CP} and in Fig. \ref{figBs3CP}
is the value of the branching ratios that, for a given pair $(\vcb,\,\vub)$, mostly deviates from the corresponding SM prediction. The resulting value of the branching fractions can be read from the legends on the right of each plot.  The benefit of these plots with respect to those already shown is that it is possible to relate a given value of the branching fractions to the entries for $(\vcb,\,\vub)$, an information that is hidden in Fig. \ref{figBsvsBd}.  The SM result as function of $(\vcb,\,\vub)$ can be read from Fig. \ref{figBsSMCP}: comparison between these plots and the corresponding one in a given 331 model would give an idea of the possible deviation as a function of $(\vcb,\,\vub)$. In particular, one can observe that M3 and M13 perform rather similarly to the SM, with values of the branching fractions that increase with $\vcb$ almost independently on $\vub$. On the other hand, this pattern is not followed in M1 and M16.


\subsection{ Rare Kaon decays}

\begin{figure}[!tb]
\begin{center}
\includegraphics[width = 0.39\textwidth]{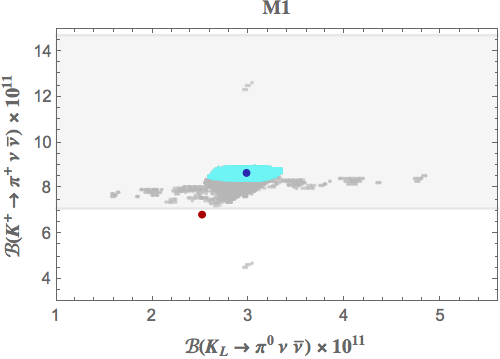}\hskip 0.5cm
\includegraphics[width = 0.54\textwidth]{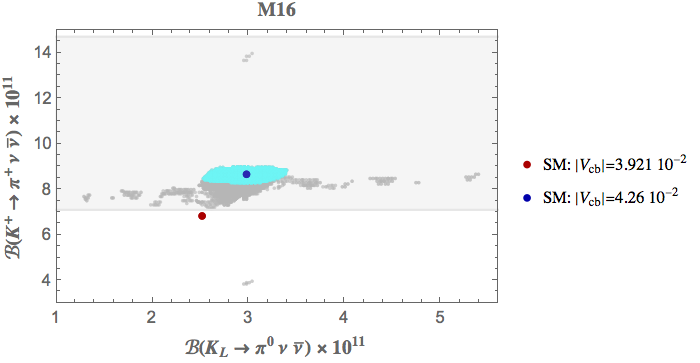}\\ \vskip 0.2cm
\includegraphics[width = 0.39\textwidth]{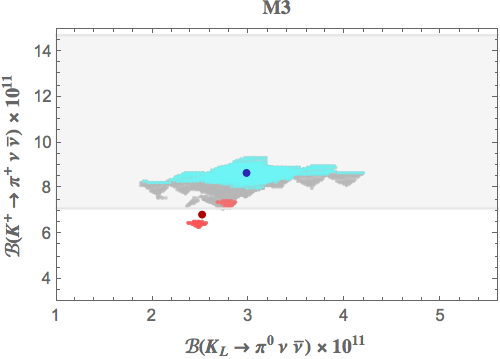}\hskip 0.5cm
\includegraphics[width = 0.54\textwidth]{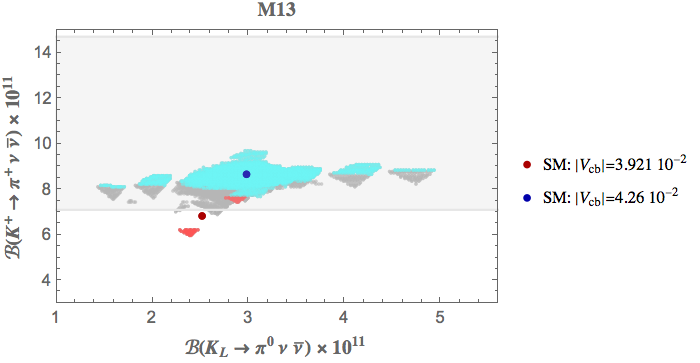}
    \caption{\small  Correlation between $ {\cal B}(K^+ \to \pi^+ \nu  \bar \nu) $ and $ {\cal B}(K_L \to \pi^0 \nu  \bar \nu) $. The gray points span all the allowed parameter space in each scenario.
    The red region corresponds to $|V_{cb}| \in [0.0386,\,0.0398]$ while the cyan region corresponds to $|V_{cb}| \in [0.0422,\,0.043]$. The SM results in correspondence of two values of $|V_{cb}|$ are displayed, as specified in the legends. The light gray region corresponds to
the experimental range for ${\cal B}(K^+ \to \pi^+ \nu  \bar \nu) $ reported in Table 1.}\label{figKpiuvsKL}
\end{center}
\end{figure}

\begin{figure}[!tb]
\begin{center}
\includegraphics[width = 0.45\textwidth]{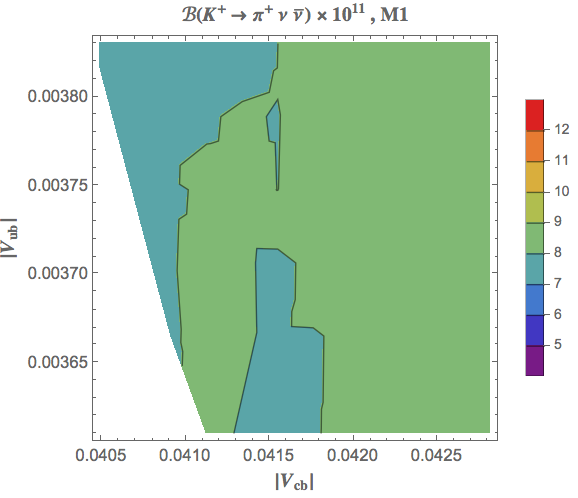}\hskip 0.5cm
\includegraphics[width = 0.45\textwidth]{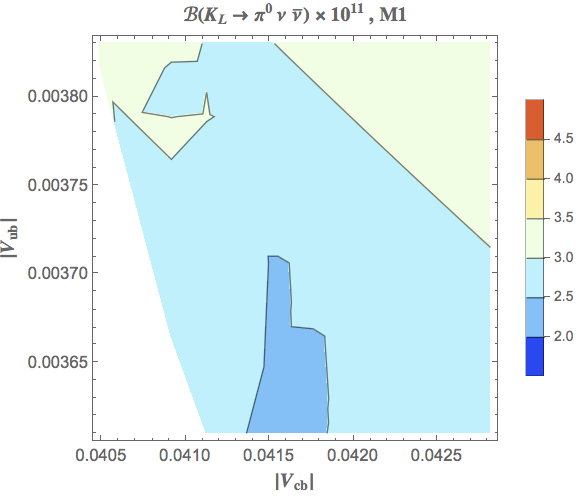} \\ \vskip 0.2cm
\includegraphics[width = 0.45\textwidth]{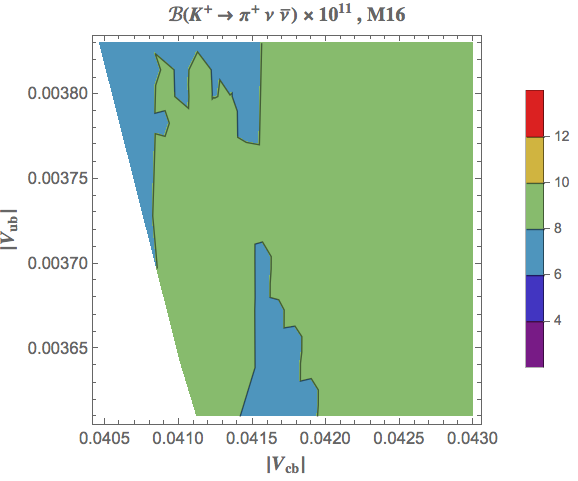}\hskip 0.5cm
\includegraphics[width = 0.45\textwidth]{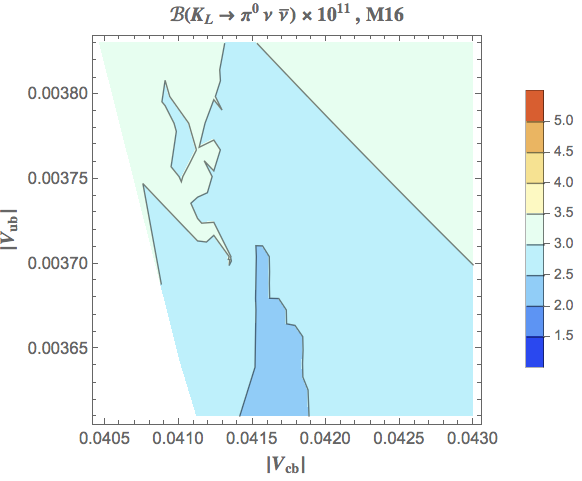}
    \caption{\small  Contour Plots of $ {\cal B}(K^+ \to \pi^+ \nu  \bar \nu) $ (left column) and $ {\cal B}(K_L \to \pi^0 \nu  \bar \nu) $ (right column) versus $|V_{cb}|$ and $|V_{ub}|$ in M1 (upper plots) and in M16 (lower plots). }\label{figK1CP}
\end{center}
\end{figure}

\begin{figure}[!tb]
\begin{center}
\includegraphics[width = 0.45\textwidth]{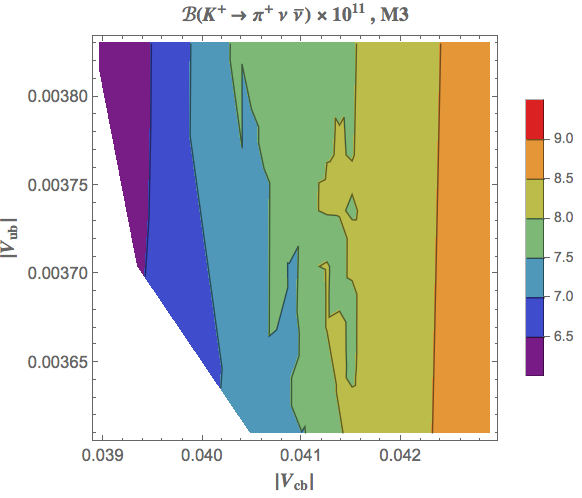}\hskip 0.5cm
\includegraphics[width = 0.45\textwidth]{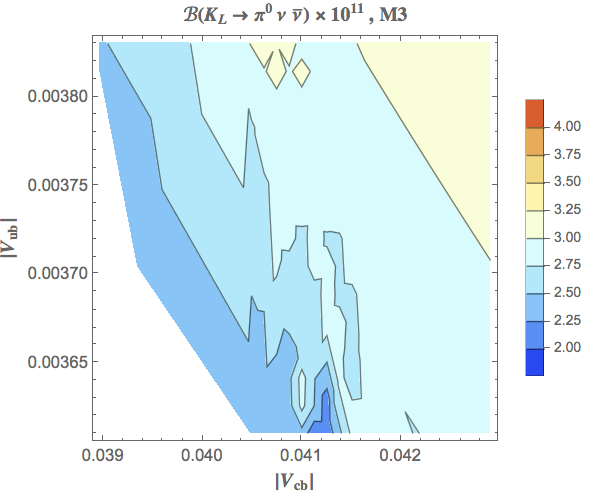}\\ \vskip 0.2cm
\includegraphics[width = 0.45\textwidth]{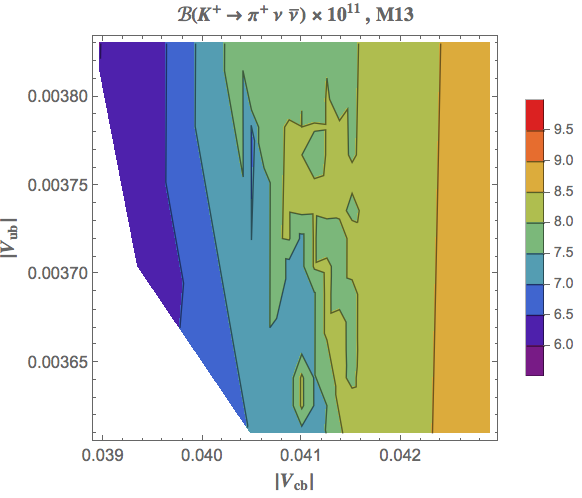}\hskip 0.5cm
\includegraphics[width = 0.45\textwidth]{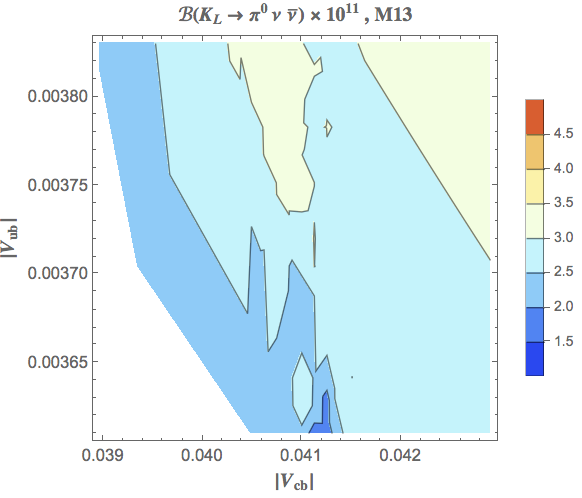}
    \caption{\small  Contour Plots of $ {\cal B}(K^+ \to \pi^+ \nu  \bar \nu) $ (left column) and $ {\cal B}(K_L \to \pi^0 \nu  \bar \nu) $ (right column) versus $|V_{cb}|$ and $|V_{ub}|$ in M3 (upper plots)and  in M13  (lower plots). }\label{figK3CP}
\end{center}
\end{figure}

\begin{figure}[!tb]
\begin{center}
\includegraphics[width = 0.45\textwidth]{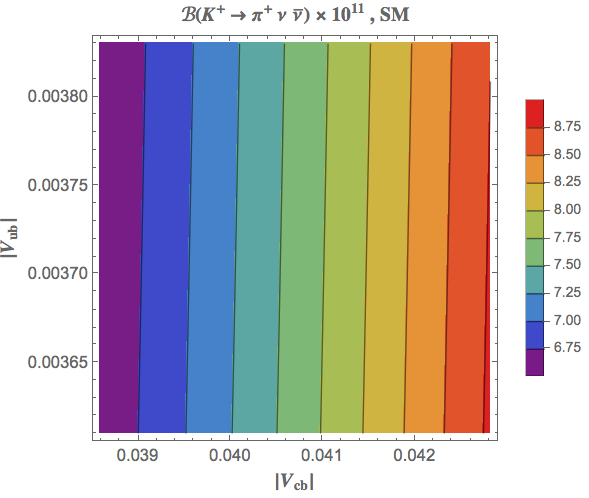}\hskip 0.5cm
\includegraphics[width = 0.45\textwidth]{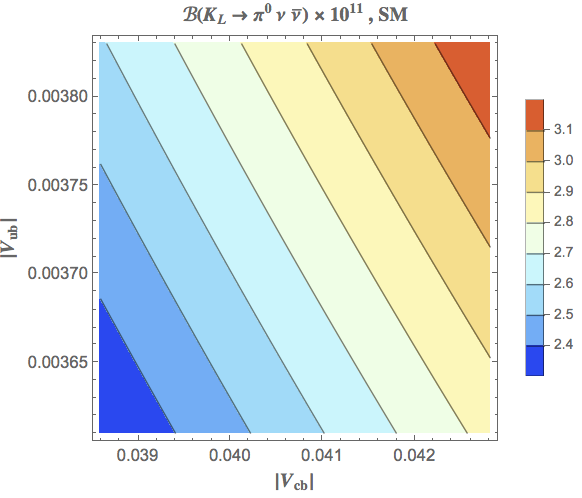} 
    \caption{\small  Contour Plots of $ {\cal B}(K^+ \to \pi^+ \nu  \bar \nu) $ (left column) and $ {\cal B}(K_L \to \pi^0 \nu  \bar \nu) $  (right column) versus $|V_{cb}|$ and $|V_{ub}|$ in the  SM . }\label{figKaonSMCP}
\end{center}
\end{figure}

In Fig. \ref{figKpiuvsKL} we display the correlation between $ {\cal B}(K^+ \to \pi^+ \nu  \bar \nu) $ and $ {\cal B}(K_L \to \pi^0 \nu  \bar \nu) $. The gray points span all the allowed parameter space in each scenario, while
the red region corresponds to $|V_{cb}| \in [0.0386,\,0.0398]$ and the cyan region  to $|V_{cb}| \in [0.0422,\,0.043]$. The SM results for $\vcb=3.921\,10^{-2}$ and $\vcb=4.26 \, 10^{-2}$ are also displayed. 
In all the four models,  the largest deviation from SM is possible in the case of $ {\cal B}(K_L \to \pi^0 \nu  \bar \nu) $.
Contour plots analogous to those presented for $B_s,\,B_d$ decays are shown in Figs. \ref{figK1CP} and \ref{figK3CP}, to be compared with the corresponding SM case in Fig. \ref{figKaonSMCP}. We observe again that M3 and M13 behave similarly to the SM, while M1 and M16 show a different pattern.

\begin{figure}[!tb]
\begin{center}
\includegraphics[width = 0.39\textwidth]{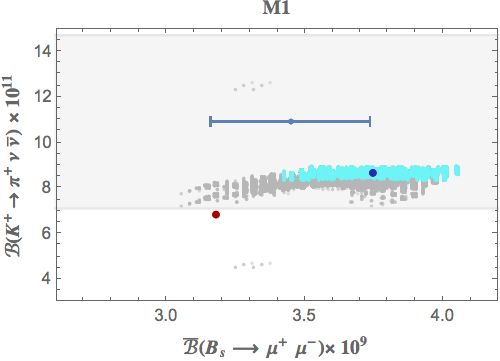}\hskip 0.5cm
\includegraphics[width = 0.54\textwidth]{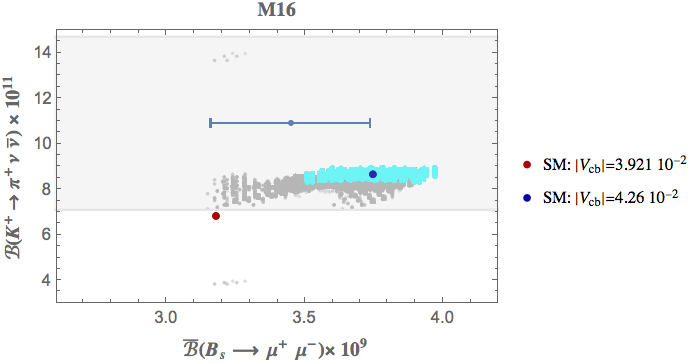}\\ \vskip 0.2cm
\includegraphics[width = 0.39\textwidth]{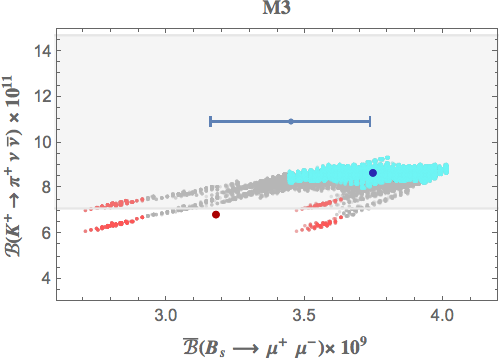}\hskip 0.5cm
\includegraphics[width = 0.54\textwidth]{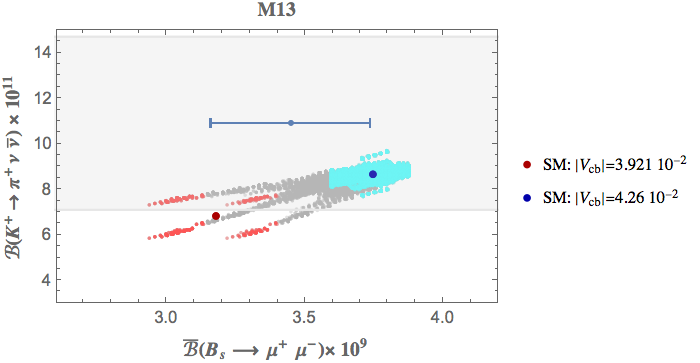}
    \caption{\small  Correlation between $ {\cal B}(K^+ \to \pi^+ \nu  \bar \nu) $ and ${\bar {\cal B}}(B_s \to \mu^+ \mu^-) $. The gray points span all the allowed parameter space in each scenario.
    The red region corresponds to $|V_{cb}| \in [0.0386,\,0.0398]$ while the cyan region corresponds to $|V_{cb}| \in [0.0422,\,0.043]$. The SM results in correspondence of two values of $|V_{cb}|$ are displayed, as specified in the legends. The light gray region and the blue range correspond to
the experimental range for ${\cal B}(K^+ \to \pi^+ \nu  \bar \nu) $ and ${\bar {\cal B}}(B_s \to \mu^+ \mu^-) $, respectively, reported in Table 1.}\label{figKpiuvsBs}
\end{center}
\end{figure}

 Correlation between $ {\cal B}(K^+ \to \pi^+ \nu  \bar \nu) $ and ${\bar {\cal B}}(B_s \to \mu^+ \mu^-) $ is shown in Fig. \ref{figKpiuvsBs}. It can be observed that in all the four cases the inclusive values of $\vcb$ correspond to points that can be compatible with the experimental result for ${\bar {\cal B}}(B_s \to \mu^+ \mu^-) $ performing slightly better than the SM; such points correspond to $ {\cal B}(K^+ \to \pi^+ \nu  \bar \nu)\le 10^{10} $.
Exclusive values of $\vcb$ that are not allowed in M1 and M16, can produce in M3 and M13 also values of ${\bar {\cal B}}(B_s \to \mu^+ \mu^-) $ and $ {\cal B}(K^+ \to \pi^+ \nu  \bar \nu) $ simultaneously smaller than the experimental range. 


\section{Summary}\label{sec:5}
Motivated by several changes both on experimental and theoretical
frontiers we updated our 2016 analysis of various flavour observables in  the
331 model   based on the gauge group $SU(3)_C\times SU(3)_L\times U(1)_X$ {for $M_{Z^\prime}=3\tev$, that is still  in the LHC reach.}

Among 24 331  models considered in our 2016 analysis only four, namely M1, M3, M13  and M16 are simultaneously consistent with the 
  electroweak precision tests  and the relation 
  between $C_9^\text{NP}$ and $C_{10}^\text{NP}$ {signalled by the most
    recent data on the $B\to \mu^+\mu^-$ decay from the CMS.}

The lessons from this analysis are as follows:

\begin{itemize}
\item
  The 331 models allow for the values of the ratio $C_9^\text{NP}/C_{10}^\text{NP}$ that are consistent with the most recent data. M13 and M16 are performing best
  but this can only be decided when new overall fits will be performed.
\item
  However, only models M1 and M16 can reach the values  ${\rm Re} [C_9^\text{NP}]=-0.7$, which although likely not  quite sufficient to explain properly the
  the suppression of $b\to s\mu^+\mu^-$ branching ratios, they reproduce
  a significant portion of it. For M3 and M13 models only the corresponding values of $-0.5$ can be reached.
\item
  {Moreover, we notice} that while in the case M1 and M16 models
  the maximal negative shifts of ${\rm Re}[C_9]$ can still be obtained for
  inclusive values in the ballpark of $\vcb=0.0415$, in the case of
  M3 and M13 the shift of $-0.5$ can only be obtained for exclusive values
  of $\vcb$ as low as $0.039$. We conclude then that models M1 and M16
  perform best in this context but as seen in Fig.~\ref{figC9Vcb}
  for the case of the HYBRID scenario for CKM parameters none of the models
  can provide suppression of ${\rm Re}[C_9]$ by more than $-0.2$ which appears
  too small from present perspective.
\item
  Concerning ${\rm Re} [C_{10}^\text{NP}]$ all models show only a small shift which
  is consistent with the data. This is also the case of of the imaginary parts
  of both $C_{9}^\text{NP}$ and $C_{10}^\text{NP}$.
\item
  As seen in Fig~\ref{figKpiuvsKL}, NP effects in $\kpn$ turn out to be
  small but could be significantly larger in $\klpn$.
\end{itemize}
{We are looking forward to improved data on all observables to be able
  to judge better the ability of the 331 models in explaining signs of
  NP.}

\section*{Acknowledgements}
{A.J.B would like to thank Andreas Crivellin for the discussion on the
present status of leptoquark models after new LHCb and CMS data.}
 This research was done in the context of the Excellence Cluster ORIGINS,
funded by the Deutsche Forschungsgemeinschaft (DFG, German Research
Foundation), 
Excellence Strategy, EXC-2094, 390783311. It 
has also been carried out within the INFN project (Iniziativa Specifica) QFT-HEP.

\bibliographystyle{JHEP}
\bibliography{bookallrefs}

\providecommand{\href}[2]{#2}\begingroup\raggedright\begin{thebibliography}{10}

\bibitem{Buras:2020xsm}
A.~J. Buras, {\em {Gauge Theory of Weak Decays}}.
\newblock Cambridge University Press, 6, 2020.

\bibitem{Pisano:1991ee}
F.~Pisano and V.~Pleitez, {\it {An SU(3) x U(1) model for electroweak
  interactions}},  {\em Phys.~Rev.} {\bf D46} (1992) 410--417,
  [\href{http://arxiv.org/abs/hep-ph/9206242}{{\tt hep-ph/9206242}}].

\bibitem{Frampton:1992wt}
P.~H. Frampton, {\it {Chiral dilepton model and the flavor question}},  {\em
  Phys.~Rev.~Lett.} {\bf 69} (1992) 2889--2891.

\bibitem{Pleitez:2021abk}
V.~Pleitez, {\it {Challenges for the 3-3-1 models}},  in {\em {5th Colombian
  Meeting on High Energy Physics}}, 12, 2021.
\newblock \href{http://arxiv.org/abs/2112.10888}{{\tt arXiv:2112.10888}}.

\bibitem{CarcamoHernandez:2022fvl}
A.~E. C\'arcamo~Hern\'andez, L.~Duarte, A.~S. de~Jesus, S.~Kovalenko, F.~S.
  Queiroz, C.~Siqueira, Y.~M. Oviedo-Torres, and Y.~Villamizar, {\it {When
  Flavor Changing Interactions Meet Hadron Colliders}},
  \href{http://arxiv.org/abs/2208.08462}{{\tt arXiv:2208.08462}}.

\bibitem{Buras:2022wpw}
A.~J. Buras and E.~Venturini, {\it {The exclusive vision of rare K and B decays
  and of the quark mixing in the standard model}},  {\em Eur. Phys. J. C} {\bf
  82} (2022), no.~7 615, [\href{http://arxiv.org/abs/2203.11960}{{\tt
  arXiv:2203.11960}}].

\bibitem{Dowdall:2019bea}
R.~J. Dowdall, C.~T.~H. Davies, R.~R. Horgan, G.~P. Lepage, C.~J. Monahan,
  J.~Shigemitsu, and M.~Wingate, {\it {Neutral $B$-meson mixing from full
  lattice QCD at the physical point}},  {\em Phys. Rev. D} {\bf 100} (2019),
  no.~9 094508, [\href{http://arxiv.org/abs/1907.01025}{{\tt
  arXiv:1907.01025}}].

\bibitem{Buras:2016dxz}
A.~J. Buras and F.~De~Fazio, {\it {331 Models Facing the Tensions in $\Delta
  F=2$ Processes with the Impact on $\varepsilon^\prime/\varepsilon$,
  $B_s\to\mu^+\mu^-$ and $B\to K^*\mu^+\mu^-$}},  {\em JHEP} {\bf 08} (2016)
  115, [\href{http://arxiv.org/abs/1604.02344}{{\tt arXiv:1604.02344}}].

\bibitem{Blanke:2016bhf}
M.~Blanke and A.~J. Buras, {\it {Universal Unitarity Triangle 2016 and the
  tension between $\Delta M_{s,d}$ and $\varepsilon _K$ in CMFV models}},  {\em
  Eur. Phys. J.} {\bf C76} (2016), no.~4 197,
  [\href{http://arxiv.org/abs/1602.04020}{{\tt arXiv:1602.04020}}].

\bibitem{Gubernari:2022hxn}
N.~Gubernari, M.~Reboud, D.~van Dyk, and J.~Virto, {\it {Improved theory
  predictions and global analysis of exclusive $b \to s\mu^+\mu^-$ processes}},
   {\em JHEP} {\bf 09} (2022) 133, [\href{http://arxiv.org/abs/2206.03797}{{\tt
  arXiv:2206.03797}}].

\bibitem{LHCb:2022qnv}
{\bf LHCb} Collaboration, {\it {Test of lepton universality in $b \rightarrow s
  \ell^+ \ell^-$ decays}},  \href{http://arxiv.org/abs/2212.09152}{{\tt
  arXiv:2212.09152}}.

\bibitem{LHCb:2022zom}
{\bf LHCb} Collaboration, {\it {Measurement of lepton universality parameters
  in $B^+\to K^+\ell^+\ell^-$ and $B^0\to K^{*0}\ell^+\ell^-$ decays}},
  \href{http://arxiv.org/abs/2212.09153}{{\tt arXiv:2212.09153}}.

\bibitem{LHCb:2021dcr}
{\bf LHCb} Collaboration, R.~Aaij et~al., {\it {Simultaneous determination of
  CKM angle $\gamma$ and charm mixing parameters}},  {\em JHEP} {\bf 12} (2021)
  141, [\href{http://arxiv.org/abs/2110.02350}{{\tt arXiv:2110.02350}}].

\bibitem{Buras:2021nns}
A.~J. Buras and E.~Venturini, {\it {Searching for New Physics in Rare $K$ and
  $B$ Decays without $|V_{cb}|$ and $|V_{ub}|$ Uncertainties}},  {\em Acta
  Phys. Polon. B} {\bf 53} (9, 2021) A1,
  [\href{http://arxiv.org/abs/2109.11032}{{\tt arXiv:2109.11032}}].

\bibitem{Buras:2022qip}
A.~J. Buras, {\it {Standard Model Predictions for Rare K and B Decays without
  New Physics Infection}},  \href{http://arxiv.org/abs/2209.03968}{{\tt
  arXiv:2209.03968}}.

\bibitem{Bordone:2021oof}
M.~Bordone, B.~Capdevila, and P.~Gambino, {\it {Three loop calculations and
  inclusive $\vcb$}},  {\em Phys. Lett. B} {\bf 822} (2021) 136679,
  [\href{http://arxiv.org/abs/2107.00604}{{\tt arXiv:2107.00604}}].

\bibitem{NA62:2022hqi}
{\bf NA62} Collaboration, M.~Zamkovsk\'y et~al., {\it {Measurement of the very
  rare $K^+ \to \pi^+ \nu \bar\nu$ decay}},  {\em PoS} {\bf DISCRETE2020-2021}
  (2022) 070.

\bibitem{Ahn:2018mvc}
{\bf KOTO} Collaboration, J.~Ahn et~al., {\it {Search for the $K_L \!\to\!
  \pi^0 \nu \overline{\nu}$ and $K_L \!\to\! \pi^0 X^0$ decays at the J-PARC
  KOTO experiment}},  {\em Phys. Rev. Lett.} {\bf 122} (2019), no.~2 021802,
  [\href{http://arxiv.org/abs/1810.09655}{{\tt arXiv:1810.09655}}].

\bibitem{Aaij:2017tia}
{\bf LHCb} Collaboration, R.~Aaij et~al., {\it {Improved limit on the branching
  fraction of the rare decay ${{K} ^0_{\mathrm { \scriptscriptstyle S}}}
  \rightarrow \mu ^+\mu ^-$}},  {\em Eur. Phys. J.} {\bf C77} (2017), no.~10
  678, [\href{http://arxiv.org/abs/1706.00758}{{\tt arXiv:1706.00758}}].

\bibitem{LHCb:2021awg}
{\bf LHCb} Collaboration, R.~Aaij et~al., {\it {Measurement of the
  $B^0_s\to\mu^+\mu^-$ decay properties and search for the $B^0\to\mu^+\mu^-$
  and $B^0_s\to\mu^+\mu^-\gamma$ decays}},
  \href{http://arxiv.org/abs/2108.09283}{{\tt arXiv:2108.09283}}.

\bibitem{CMS:2020rox}
{\bf CMS} Collaboration, {\it {Combination of the ATLAS, CMS and LHCb results
  on the $B^0_{(s)} \to \mu^+\mu^-$ decays}},
  \href{http://arxiv.org/abs/CMS-PAS-BPH-20-003}{{\tt CMS-PAS-BPH-20-003}}.

\bibitem{ATLAS:2020acx}
{\bf ATLAS} Collaboration, {\it {Combination of the ATLAS, CMS and LHCb results
  on the $B^0_{(s)}\to\mu^+\mu^-$ decays.}},
  \href{http://arxiv.org/abs/ATLAS-CONF-2020-049}{{\tt ATLAS-CONF-2020-049}}.

\bibitem{HFLAV:2022pwe}
{\bf HFLAV} Collaboration, Y.~Amhis et~al., {\it {Averages of $b$-hadron,
  $c$-hadron, and $\tau$-lepton properties as of 2021}},
  \href{http://arxiv.org/abs/2206.07501}{{\tt arXiv:2206.07501}}.

\bibitem{Zyla:2020zbs}
{\bf Particle Data Group} Collaboration, P.~A. Zyla et~al., {\it {Review of
  Particle Physics}},  {\em PTEP} {\bf 2020} (2020), no.~8 083C01.

\bibitem{Buras:2014yna}
A.~J. Buras, F.~De~Fazio, and J.~Girrbach-Noe, {\it {Z-Z' mixing and Z-mediated
  FCNCs in $SU(3)_C \times SU(3)_L \times U(1)_X$ Models}},  {\em JHEP} {\bf
  1408} (2014) 039, [\href{http://arxiv.org/abs/1405.3850}{{\tt
  arXiv:1405.3850}}].

\bibitem{Buras:2012dp}
A.~J. Buras, F.~De~Fazio, J.~Girrbach, and M.~V. Carlucci, {\it {The Anatomy of
  Quark Flavour Observables in 331 Models in the Flavour Precision Era}},  {\em
  JHEP} {\bf 1302} (2013) 023, [\href{http://arxiv.org/abs/1211.1237}{{\tt
  arXiv:1211.1237}}].

\bibitem{Buras:2013dea}
A.~J. Buras, F.~De~Fazio, and J.~Girrbach, {\it {331 models facing new $b \to
  s\mu^+ \mu^-$ data}},  {\em JHEP} {\bf 1402} (2014) 112,
  [\href{http://arxiv.org/abs/1311.6729}{{\tt arXiv:1311.6729}}].

\bibitem{Buras:2015kwd}
A.~J. Buras and F.~De~Fazio, {\it {$\varepsilon'/\varepsilon$ in 331 Models}},
  {\em JHEP} {\bf 03} (2016) 010, [\href{http://arxiv.org/abs/1512.02869}{{\tt
  arXiv:1512.02869}}].

\bibitem{Colangelo:2021myn}
P.~Colangelo, F.~De~Fazio, and F.~Loparco, {\it {$c\to u {\bar \nu}{\nu}$
  transitions of Bc mesons: 331 model facing Standard Model null tests}},  {\em
  Phys. Rev. D} {\bf 104} (2021), no.~11 115024,
  [\href{http://arxiv.org/abs/2107.07291}{{\tt arXiv:2107.07291}}].

\bibitem{Buras:2021rdg}
A.~J. Buras, P.~Colangelo, F.~De~Fazio, and F.~Loparco, {\it {The charm of
  331}},  {\em JHEP} {\bf 10} (2021) 021,
  [\href{http://arxiv.org/abs/2107.10866}{{\tt arXiv:2107.10866}}].

\bibitem{Aoki:2019cca}
{\bf Flavour Lattice Averaging Group} Collaboration, S.~Aoki et~al., {\it {FLAG
  Review 2019: Flavour Lattice Averaging Group (FLAG)}},  {\em Eur. Phys. J. C}
  {\bf 80} (2020), no.~2 113, [\href{http://arxiv.org/abs/1902.08191}{{\tt
  arXiv:1902.08191}}].

\bibitem{Aoki:2021kgd}
Y.~Aoki et~al., {\it {FLAG Review 2021}},
  \href{http://arxiv.org/abs/2111.09849}{{\tt arXiv:2111.09849}}.

\bibitem{Brod:2021hsj}
J.~Brod, M.~Gorbahn, and E.~Stamou, {\it {Updated Standard Model Prediction for
  $K \to \pi \nu \bar{\nu}$ and $\epsilon_K$}},  in {\em {19th International
  Conference on B-Physics at Frontier Machines}}, 5, 2021.
\newblock \href{http://arxiv.org/abs/2105.02868}{{\tt arXiv:2105.02868}}.

\bibitem{Brod:2019rzc}
J.~Brod, M.~Gorbahn, and E.~Stamou, {\it {Standard-Model Prediction of
  $\epsilon_K$ with Manifest Quark-Mixing Unitarity}},  {\em Phys. Rev. Lett.}
  {\bf 125} (2020), no.~17 171803, [\href{http://arxiv.org/abs/1911.06822}{{\tt
  arXiv:1911.06822}}].

\bibitem{Buras:2010pza}
A.~J. Buras, D.~Guadagnoli, and G.~Isidori, {\it {On $\epsilon_K$ beyond lowest
  order in the Operator Product Expansion}},  {\em Phys.~Lett.} {\bf B688}
  (2010) 309--313, [\href{http://arxiv.org/abs/1002.3612}{{\tt
  arXiv:1002.3612}}].

\bibitem{Buras:1990fn}
A.~J. Buras, M.~Jamin, and P.~H. Weisz, {\it {Leading and next-to-leading QCD
  corrections to $\varepsilon$ parameter and $B^0-\bar{B}^0$ mixing in the
  presence of a heavy top quark}},  {\em Nucl.~Phys.} {\bf B347} (1990)
  491--536.

\bibitem{Urban:1997gw}
J.~Urban, F.~Krauss, U.~Jentschura, and G.~Soff, {\it {Next-to-leading order
  QCD corrections for the $B^0 - \bar B^0$ mixing with an extended Higgs
  sector}},  {\em Nucl.~Phys.} {\bf B523} (1998) 40--58,
  [\href{http://arxiv.org/abs/hep-ph/9710245}{{\tt hep-ph/9710245}}].

\bibitem{Amhis:2016xyh}
{\bf Heavy Flavor Averaging Group (HFAG)} Collaboration, Y.~Amhis et~al., {\it
  {Averages of $b$-hadron, $c$-hadron, and $\tau$-lepton properties as of
  summer 2016}},  \href{http://arxiv.org/abs/1612.07233}{{\tt
  arXiv:1612.07233}}.

\end{thebibliography}\endgroup
\end{document}